%% file: ASCOM_2016_87_revision.tex
\documentclass[review]{elsarticle}
\usepackage{lineno,hyperref}
\modulolinenumbers[5]

\newcommand{\nb}{\textit{N}--body }

\usepackage[]{algorithmic}

\begin{document}

\begin{frontmatter}

\title{Massively Parallel Computation of Accurate Densities for \nb Dark Matter Simulations using the Phase-Space-Element Method}

\author{Ralf Kaehler}
\address{Kavli Institute for Particle Astrophysics and Cosmology,\\
SLAC  National Accelerator Laboratory \\
Menlo Park, CA  94025, USA}



\input{abstract.tex}


\end{frontmatter}


\input{main.tex}

\bibliography{mybibfile}

\end{document}

%% file: abstract.tex
\begin{abstract}
This paper presents an accurate density computation
approach for large dark matter simulations, based on a recently introduced phase-space
tessellation technique and designed for massively parallel, heterogeneous cluster
architectures. 

We discuss a memory efficient construction of an oct-tree structure to
sample the mass densities with locally adaptive resolution,
according to the features of the underlying tetrahedral tessellation.
We propose an efficient GPU implementation for the computationally intensive
operation of intersecting the tetrahedra with the cubical cells of the deposit grid,
that achieves a speedup of almost an order of magnitude compared to an
optimized CPU version. We discuss two dynamic load balancing schemes - the first 
exchanges particle data between cluster nodes and deposits all
tetrahedra for each block of the grid structure on single nodes,
whereas the second approach uses global reduction operations to obtain the total masses.
We demonstrate the scalability of our algorithms for up to $256$ GPUs and
TB--sized simulation snapshots, resulting in tessellations with over $400$ billion tetrahedra. 
\end{abstract}

%% file: main.tex
\section{Motivation}
Dark matter is a key component in state--of--the--art large--scale
structure formation theories, and \nb simulations have become an
essential method to test their predictions.
This numerical approach treats dark matter as a collisionless gas, that is sampled by a set of 
particles of equal mass, whose positions are updated over time, according to the overall gravitational
forces
\cite{Dikaiakos:1996:PSC:237578.237590, 0067-0049-111-1-73,
  Teyssier:2001cp, Dubinski:2004, Springel05thecosmological,
  0067-0049-184-2-298, doi:10.1093/pasj/61.6.1319,
  Habib:2013:HES:2503210.2504566, Warren:2013:IPH:2503210.2503220,
  2014ApJS..211...19B, 0004-637X-765-1-39}.


An integral part of \nb simulations, as well as applications that rely on an
analysis of the resulting datasets, is the computation of accurate
mass densities, for example, to solve the Poisson equation, identify features of the cosmic web or to predict
dark matter annihilation signals. There exist various techniques to extrapolate from the mass at the discrete particle positions to
a density field defined everywhere in the computational domain.
{\sl Particles Mesh} codes, for example, construct the underlying density field 
at the vertices, respectively cells of an auxiliary grid structure. 
The simplest technique, also known as {\sl Nearest Grid Point
  (NGP)}~\cite{Hockney:1988:CSU:62815}, 
assigns the mass associated with each particle to the grid cell that contains it.
The {\sl Cloud-In-Cell}~\cite{1969JCoPh...3..494B} approach models the particle's mass distribution as a cube
centered at its position and distributes the mass proportionally to all
overlapping cells of a regular grid. {\sl Smoothed-Particle Hydrodynamics (SPH)}~schemes superimpose rotationally
symmetric kernel functions centered at the particles~\cite{1988CoPhC..48...89M}. 
Another technique is to construct Voronoi tessellations from the particles'
positions, estimating the densities using the particles' enclosing volumes~\cite{2008MNRAS.386.2101N}. 
However, all these methods are subject to artifacts due to sampling noise inherent to the underlying discrete
distributions, which is problematic in many applications.

In 2012, a density computation method for \nb
simulations, not affected by the above-mentioned Poisson noise, 
was introduced~\cite{Shandarin:2011jv, 2011AbelHahnKaehler}.
The basic idea is to use the tracer particles to construct a tessellation of
the 3-dimensional dark matter sheet, which is embedded in a
6-dimensional phase space. Hereby the mass is spread out across the
elements of the tessellation. Projecting the tessellation into
configuration space and adding the density contributions from all
elements that overlap the same spatial location, gives rise to well--defined 
densities in the entire computational domain.

This method has been successfully
applied to reduce artificial clumping in N­-body
simulations~\cite{Hahn:2012ma}, 
to improve the quality of visualizations of
dark matter simulations~\cite{10.1109..TVCG.2012.187,Igouchkine:2016:VRD:3002151.3002163},
to create smooth maps of the gravitational lensing
potential around dark matter halos~\cite{Angulo:2013bfq}, and to study the
statistics of cosmic velocity fields~\cite{Hahn:2014lca}. It also led
to the construction of new numerical schemes for dark matter
simulations~\cite{Hahn:2015sia}, and inspired work in 
computational geometry~\cite{Powell:2014hea}.

A drawback of the phase-space tessellation approach is its
computational complexity, resulting from the large amount of
tessellation elements that need to be processed.  
Fortunately, the inherently parallel nature of the problem is
well--suited for massively parallel, heterogeneous cluster 
architectures equipped with accelerators, which recently have become very
popular, because of their good performance and energy efficiency.

This paper presents a phase-space tessellation-based 
density computation approach for large \nb dark matter simulation data,
utilizing massively parallel (GPU-)clusters. We introduce an exact
tetrahedron--cube intersection algorithm, with optimized CPU and GPU variants, to sample
the mass associated with the tetrahedral elements onto a block-structured grid. 

The resolution of the grid is locally refined, based on the
features of the underlying tessellation. Crucial for the
overall performance is an efficient load balancing scheme, that
distributes the workload equally across the cluster, while
minimizing data transfers and memory requirements. 
We further compare two dynamic
load-balancing strategies - one that gathers all tetrahedra required for
depositing the mass onto each oct-tree patch on a single cluster node,
as opposed to deposting only local tetrahedra,
followed by a reduction step to obtain the total masses. 
We end the paper with a detailed analysis of the algorithm for various datasets of up
to, including several weak and strong scaling tests.

\section{Review of the Phase-Space Element Approach}
\label{Sec-PhysicalMotivation}

\begin{figure}[htb]
  \centering
  \includegraphics[width=1.\linewidth]{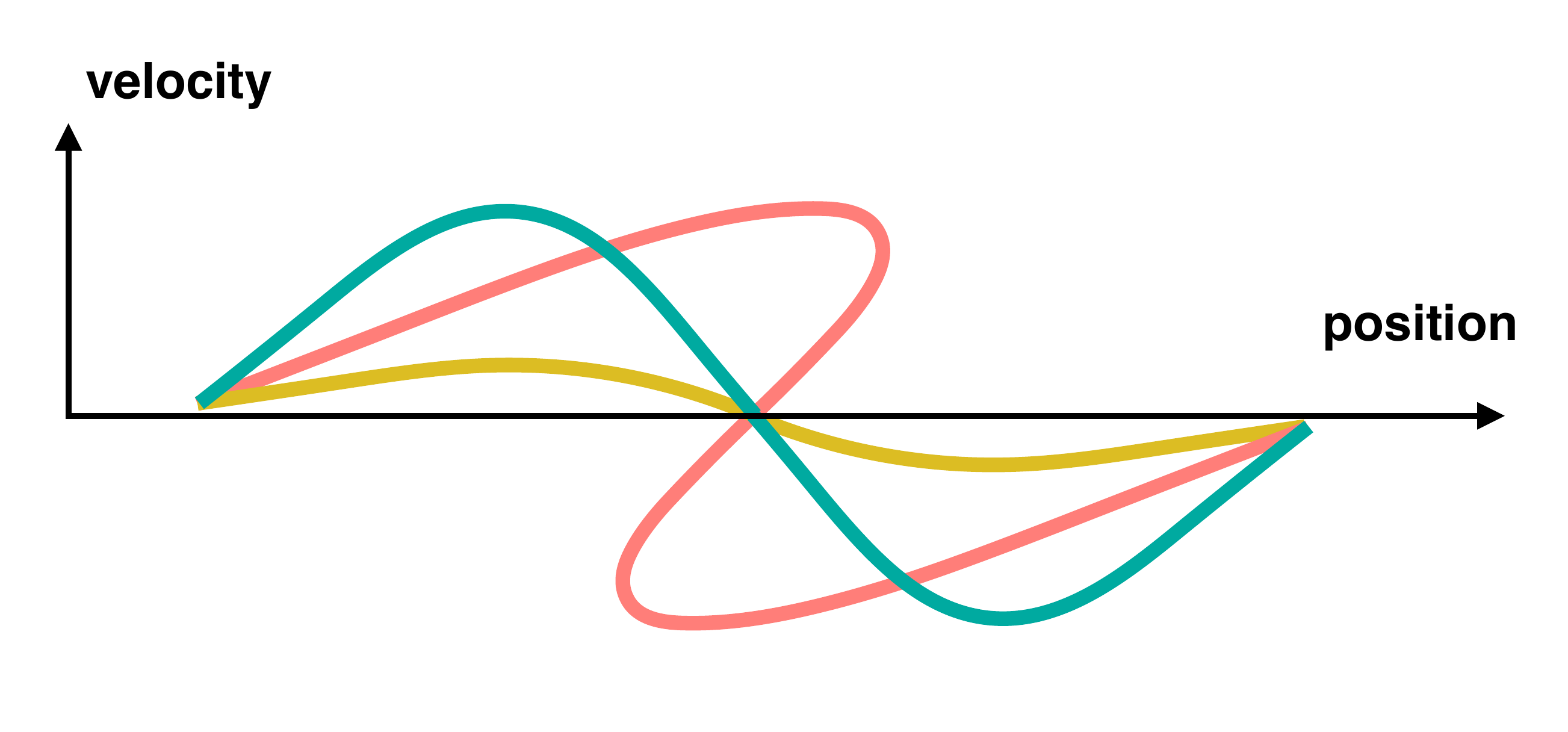}
  \caption
  {
    \label{Fig-Phase-Space}
    {
      This 2D phase-space diagram shows the positions and
      velocities of the dark matter fluid for three different
      time-steps. Initially, the dark matter is almost
      uniformly distributed and at rest (yellow). Gravity accelerates 
      the fluid elements which gain
      velocity (green) and at later times, several
      streams of the dark matter co-exist at the same spatial regions (red).
    }
  }
\end{figure}

In this section, we briefly review the main ideas of the phase-space
tessellation method, that are most relevant for this paper -- for a detailed 
discussion please refer to~\cite{Shandarin:2011jv,
  2011AbelHahnKaehler}.
\nb simulations model the large-scale evolution of 
dark matter in the Universe using point-like mass sources, called {\sl
tracer particles} or just {\sl tracers} in the following discussion.
Their positions are updated according
to the aggregate gravitational forces, under the assumption, that the 
mass is centered at the tracers' positions. However, it is physically more accurate to regard dark matter as
a collisionless fluid, governed by the {\sl Vlasov-Poisson equation}~\cite{Peebles1993}, 
with the mass spread out across the computational domain, instead of being concentrated at a number
of quite arbitrary discrete sampling locations, as illustrated in the
2-dimensional phase-space diagram in
Figure~\ref{Fig-Phase-Space}.

At the initial time step, the tracers are distributed 
almost uniformly throughout the computational domain, by
aligning their positions with the vertices of a regular grid. 
We will refer to this arrangement as the {\sl Lagrangian grid}
in what follows.
\begin{figure}[htb]
  \centering
  \mbox{} \hfill
 \includegraphics[width=0.975\linewidth]{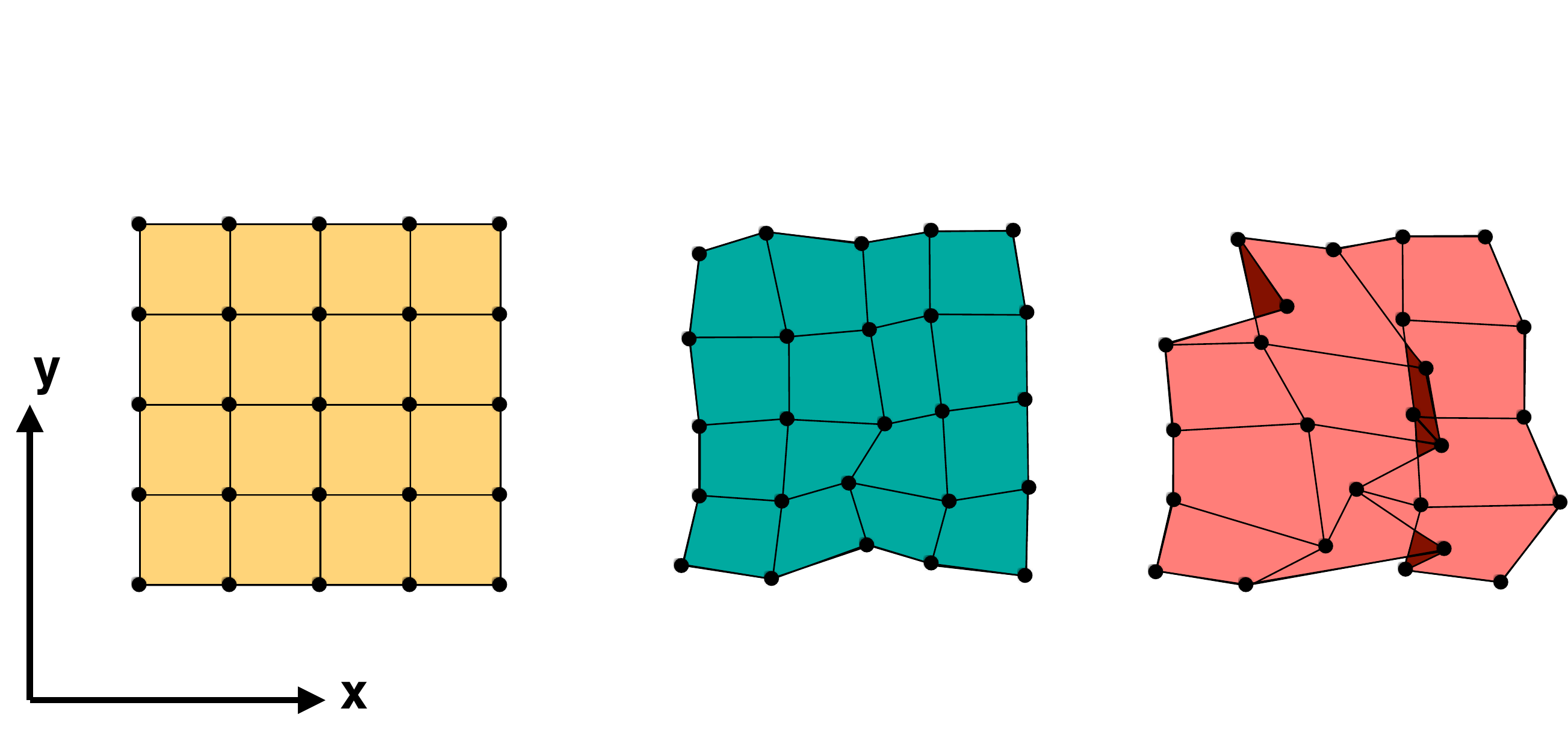}
 \hfill \mbox{}
  \caption{\label{Fig-2DMesh}%
           A 2D regular grid structure defined by
           the tracer particles of a \nb simulation. Initially, the particles are
           distributed regularly over the computational domain
           (left). Over time, the particles are advected due to
           gravitational forces and the cells become
           deformed (middle). At later times, cells finally start to 
           overlap (dark shaded regions on the right). }
\end{figure}
Since each cell initially has the same volume and that the mass is
distributed uniformly across the computational domain,
the same mass is assigned to each cell. 
The cell connectivity is fixed over time, whereas 
the spatial locations of the vertices are updated 
according to the actual positions of the tracer particles, 
causing a deformation of the embedding of the logical rectangular Lagrangian 
grid in configuration space, as depicted in Figure~\ref{Fig-2DMesh}.  
Given the constant mass per cell, its time-dependent volume
provides an estimation of its local density. However, instead of directly using the
hexahedral cells to compute these volumes, it is computational more
efficient to first tessellate them by tetrahedra, as these are always convex,
independently of the relative positions of their vertices. We will follow 
the tessellation choice using $6$ tetrahedra, as for example used in~\cite{10.1109..TVCG.2012.187}.
So if $m$ is the constant mass per tetrahedron and $\sum_{i} V_i(\mathbf{x},t)$ the 
total volume of all tetrahedra that cover the location
$\mathbf{x}$ at time $t$, the total mass density is given by
\begin{eqnarray}
\label{Eqn-rho_tot_a}
 \rho_{tot}(\mathbf{x} ,t) = m \sum_i \frac{1}{ V_i(\mathbf{x},t) }.
\end{eqnarray}

\section{Distributed Tessellation Construction}
\label{Section::DataInput}
\begin{figure*}
  \centering
\includegraphics[width=0.75\textwidth]{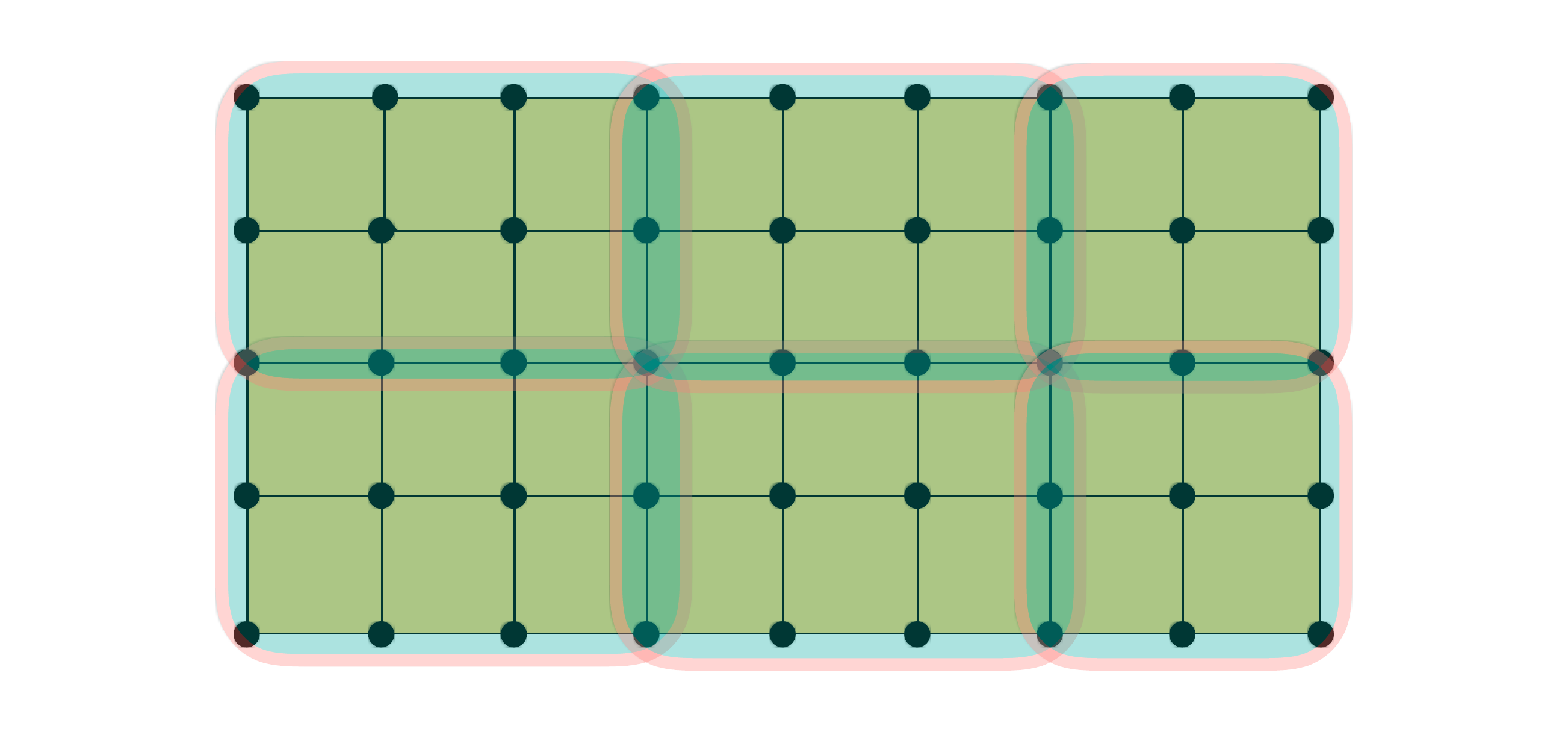}
\caption{ Example of the subdivision of a logical rectangular 2D Lagrangian grid with
  $9 \times 5$ particles into 6 blocks, that are assigned to different
  cluster nodes. One layer of particles between adjacent blocks is duplicated to
  guarantee, that all tetrahedra can be reconstructed locally. }
\label{Fig-LagrangianPatches}
\end{figure*}

In this Section, we discuss the construction of the tetrahedral
tessellation in a distributed environment. The goal is to assign approximately
the same number of elements to each node, while exploiting
as much spatial locality as possible, in order to reuse tracers that
are shared between multiple elements.
Our strategy it to subdivide the underlying logical Lagrangian grid into
rectangular blocks, and assign them to different cluster nodes.
To determine the three-dimensional layout of the blocks, 
we compute all possible factorizations of the number of nodes by 
three integers and choose the one that yields the most similar number 
of blocks along all dimensions. Tracer particles on common faces,
edges and vertices between adjacent blocks are duplicated to ensure that each
individual node has access to all data required to locally construct the tetrahedra
associated with the cells of its block. The subdivision process is
illustrated in Figure~\ref{Fig-LagrangianPatches}.

Large-scale \nb simulation codes usually store their data
in spatially clustered arrangments, for example employing space-filling 
Hilbert-curves~\cite{Springel05thecosmological}
or Morton-order~layouts~\cite{Warren:2013:IPH:2503210.2503220}. So  
that the data is already partially organized in
blocks in Lagrangian grid space, in particular for the earlier
time-steps, and since the tracers typically travel a relatively
small fraction of the overall computational domain, the correlation
between chunks of data on disk and blocks in Lagrangian
grid space is partly conserved over time. 
So loading a chunk of contiguous data from
file on each cluster node, and choosing a file offset that corresponds to the node's assigned Lagrangian
block, ensures that a large fraction of the loaded particle data is part of the node's block.
However, in general, a fraction of the tracers for each block will
be unavailable, so internode communication will be required in order
to exchange the missing data. We speed up this process by organizing the local
particles into different bins, one for each block of the Lagrangian
grid, respectively cluster node. The assignment to the bins is based on
the particles' Lagrangian positions, which is given by their unique IDs. 
Due to the shared boundary layers between adjacent Lagrangian blocks,
a fraction of particles may be duplicated in up to $8$ bins. However,
while this introduces only a small memory overhead, it
avoids expensive search operations and allows to free entire
bins efficiently, when the pairwise communication between pairs of nodes has been carried out.
Once a node received all particle data for its local block, the
tracers are sorted by their IDs, which then become obsolete and are discarded.

\section{Constructing the Adaptive Grid Structure}
\label{Section::AdaptiveGridStructure}
In this section, we discuss the distributed construction of the
adaptive deposit grid structure, an oct--tree, although an extension 
to more general structures like kD--, or AMR (adaptive
mesh refinement) trees would be possible, too.

Starting at the root of the tree, which is defined by the minimal axis--aligned bounding box
containing all tracer particles in the computational domain, all
process detect the number of local tetrahedra that cover each
node of the tree and obtains the overall number via a global
reduction. If the number is above a user-defined threshold, the 
bounding box is refined by 8 children, and the process
continues recursively, until the number of elements that partially
overlap the node's region is below the threshold or a minimal
node extension is reached. The latter is necessary because the number of
tetrahedra does not always decrease on the refined regions, for
example, once it is completely contained inside the
intersection of a set of tetrahedra. The leaf nodes of the tree are
overlaid by cubical grid patches of the same number of cells and thus 
yield a sampling of the computational domain, whose local resolution is
adapted to the density of the tetrahedra. 

A drawback of the approach discussed so far is its high memory consumption. 
since in order to efficiently check the refinement criterion, one has to keep track
of sets of local tetrahedra that cover different parts of the tree, e.g. by storing
their unique IDs at each node of the tree. But the
number of tetrahedral exceeds the number of tracer particles by a
almost a factor of~$6$, and at least $64$-bit integers are necessary to
uniquely identify them, so even in the optimal case that each
element falls into exactly one leaf node and thus needs to be stored 
only once, the memory requirements would exceed the one for storing the original tracer
positions.

\begin{figure*}
  \centering
\includegraphics[width=1.\textwidth]{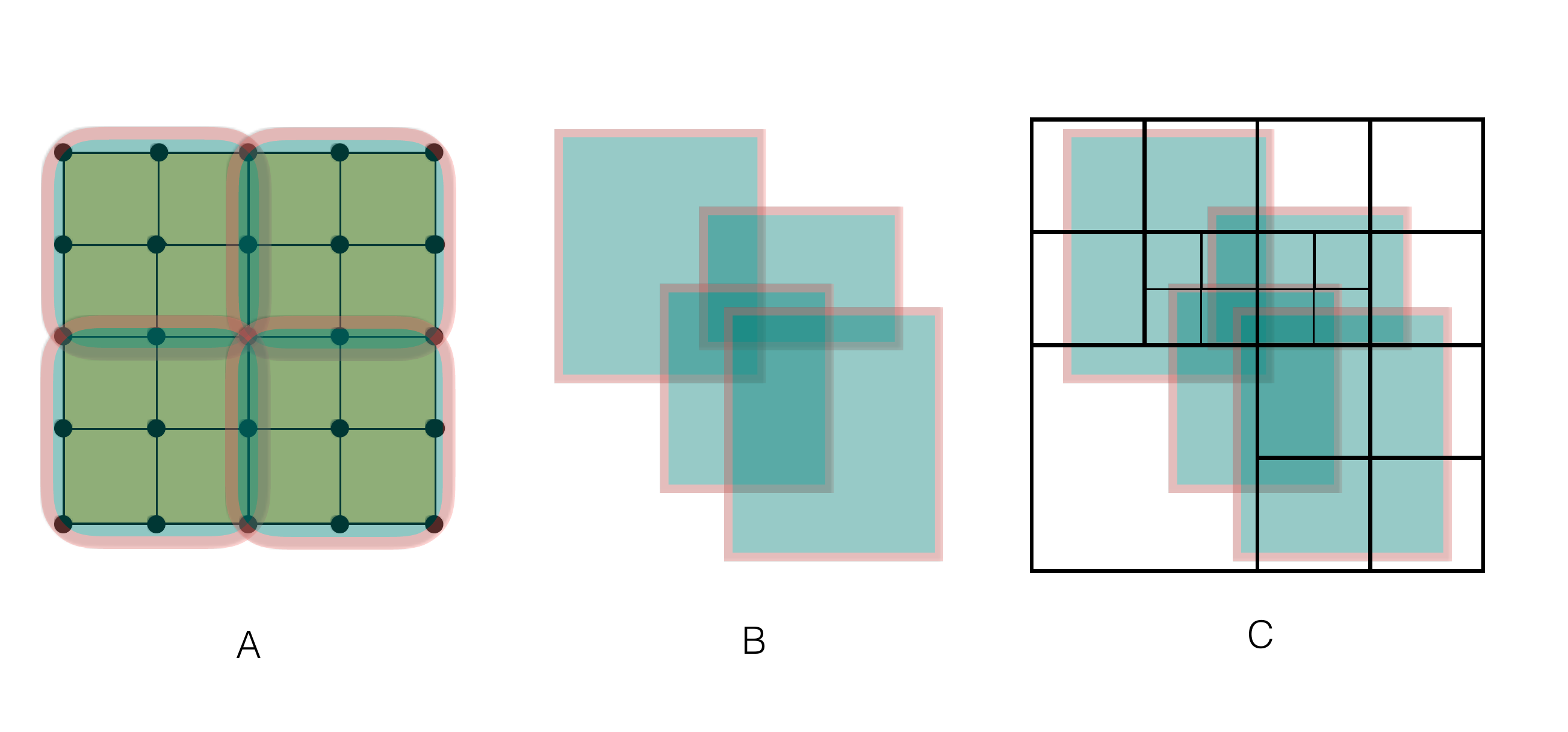}
\caption{ 2D illustration of a memory efficient oct-tree construction: (A)
  shows a Lagrangian grid block with $5 \times 5$ particles, which is 
further subdivided into 4 subblocks. (B) The minimal axis-aligned
bounding boxes, spanned by the particles of each of these subblocks, in
configuration space. (C) The quad-tree (oct-tree) structure, 
refined based on the area  (volume) of covered subblocks. }
\label{Fig-OctreeConstruction}
\end{figure*}

In order to reduce this overhead, we further partition the local Lagrangian
grid block on each node of the cluster into a set of subregions,
called subblocks in the following discussion, and precompute the partially 
overlapping, minimal axis-aligned bounding boxes enclosing the 
particles of each subblock. Instead of testing each tetrahedron
individually, we base the refinement decision on the volume of the
intersection between these bounding boxes and the tree nodes. 
The IDs of intersecting subblock 
are stored at each tree node, which drastically reduces the memory 
overhead in comparison to storing the IDs of individual tetrahedra. 
The number of tetrahedra that overlap a specific tree node is
estimated by the intersection volume between the nodes and the subblocks' bounding boxes, assuming that
the tetrahedra in each subblock have similar volumes. 
A number $8^3$ cells for each subblock worked well for our
applications, in terms of the trade-off between the number of
generated subblocks and the spatial locality of the tetrahedra.
A overview of the oct-tree construction is given in Figure~\ref{Fig-OctreeConstruction}.

\section{Dynamic Load Balancing}
\label{Sec-LoadBalancing}

In general, the tetrahedra density 
will vary substantially across the
computational domain, ranging from a few elements per
point in underdense voids, to several thousand in dark matter halos. 
To some extent, we account
for this variance by adapting the spatial resolution of the oct-tree
to the density of the underlying tessellation. However, the workload
still differs for the leaf nodes of the tree, so an efficient
load-balancing technique is crucial for
optimally utilizing the computational resources.

We employ dynamic load balancing strategies on three different
levels: between individual compute nodes, between individual CPU
threads on each compute node, as well as on the level of GPU kernel calls, launched 
by the CPU threads. The latter two will be discussed in
detail in Section~\ref{Sec-MassDeposit}, while here we focus on load balancing
the compute nodes.

In general, the tetrahedra required for depositing the mass onto a
leaf node of the tree structure will be scattered across multiple compute
nodes and we will discuss two options for processing each leaf: (A), gathering all required
tetrahedra on one compute node, which deposits the elements' masses,
or alternatively (B),
depositing only the local tetrahedra on each compute node, followed by a reduction operation
between all processes that contributed to the leaf. In the next
two paragraphs, we will describe our implementations of these
approaches using MPI, and defer a performance comparison between them to Section~\ref{Sec-Results}.
In both cases, we launch one MPI process per compute node and at least 
two CPU threads per MPI rank, in order to overlap local computation 
and internode communication.

\paragraph{Option (A): Exchanging Point Data}
In this approach, each MPI rank enters a loop and
signals all other ranks if it is ready for new work, by setting a flag
in a bitmask, one bit for each rank, that is globally reduced. 
The first rank R that is ready for new work, is assigned the leaf
node L, for which it locally stores most of the required tetrahedra. The other MPI ranks send their 
local tetrahedra that overlap L, to R, which starts depositing the
elements, as discussed in detail in Section~\ref{Sec-MassDeposit},
while, concurrently, new work is assigned to other MPI ranks.
Instead of constructing the tetrahedra of
the Lagrangian blocks before sending them to the target rank, 
each compute node sends complete $8^3$~subblocks of particles and
R constructs the tessellation elements locally, thereby drastically reducing the
amount of transferred data. The overall algorithm is outlined in the
pseudo-code in Figure~\ref{Algo-CommA}.
\begin{figure}[ht]
\begin{minipage} {.75\textwidth}
\small
  \begin{algorithmic} [1]
\STATE launch local worker threads
\STATE leafNodeList $\gets$ all octree leaf nodes
\WHILE{ leafNodeList NOT empty }

    \IF{ local worker threads finished }
         \STATE store mass of previous grid patch
    \ENDIF

    \STATE MPI: share which processes are ready for new work

    \FORALL{ processes R that are ready for new work }
        \STATE L $\gets$ FindBestOctreeLeafNode( R, leafNodeList )
        \STATE remove L from leafNodeList
        \STATE MPI: subblocks that overlap L are
      send to R
        \IF { this rank == R }
              \STATE local worker threads start mass deposit for L
        \ENDIF

    \ENDFOR

\ENDWHILE
\end{algorithmic}
\end{minipage}
\caption{Pseudo code for communication approach (A) on each MPI
  process. Lines beginning with '{\sl MPI:}' indicate points of internode
  communication. }
\label{Algo-CommA}
\end{figure}

\paragraph{ Option (B): Global Reduction}
In this approach, each MPI process enters a loop, picks the next leaf node
for which it has local tetrahedra and starts the mass deposit using
its set of worker threads, see Section~\ref{Sec-MassDeposit}.
Concurrently, the MPI ranks reduce a bit mask that indicates which of the
leaves are finished on each node. 
Once the mass deposit for a leaf has been completed on all cluster
nodes, the masses of the local mass arrays are combined to yield the resulting densities. 
This is achieved via a global reduction operation between all ranks
that contributed mass to the leaf node. The reduction's target rank is 
altered in a round-robin fashion, in order to distribute the workload.
 \begin{figure}[ht]
\begin{minipage} {.7\textwidth}
\small
  \begin{algorithmic}[1]
\STATE launch local worker threads
\STATE leafNodeList $\gets$ all octree leaf nodes
\STATE allocate buffer B (stores octree leaf masses)
\REPEAT
    \IF { local worker threads finished work item }
        \STATE B $\gets$ local mass of processed octree leaf
    \ENDIF
    \IF {$($ B NOT full $)$ AND $($ leafNodeList NOT empty $)$}
        \STATE L $\gets$ next leaf node in leafNodeList
        \STATE remove L from leafNodeList
        \STATE  local worker threads start mass deposit for L
   \ENDIF
    \STATE MPI: determine which leaves are globally finished
      \FORALL{newly finished octree leaves L}
        \IF { this rank contributed mass to L }
        \STATE R $\gets$ process that stores final grid patch mass
        \STATE MPI: total masses for L via reduction to R
        \IF{ this rank == R}
                \STATE store total masses for L
        \ENDIF
        \STATE remove L from buffer B
        \ENDIF
     \ENDFOR
\UNTIL{ all leaf node masses have been stored }
\end{algorithmic}
\end{minipage}
\caption{Pseudo code for communication approach (B), executed on
  each MPI process. Lines beginning with '{\sl MPI:}' indicate interprocess
  communication points. }
\label{Algo-CommB}
\end{figure}
In case the mass deposit is completed locally but not globally, 
the corresponding local mass array is buffered, and the next leaf
is processed, in order to avoid that the worker threads idle.
However, once the buffer's maximal storage capacity is reached, 
the rank has to postpone the deposit of new blocks, until at least one 
of its buffered leaf nodes is finished on all nodes.
The algorithm is outlined in Figure~\ref{Algo-CommB}.

\section{Mass Deposit}
\label{Sec-MassDeposit}

As discussed in the last Section, concurrent computation and
communication is enabled by launching several threads per 
MPI process: one communication thread, the main thread, for data
exchange between cluster nodes, as well as at least one worker thread
to manage the deposition of the tetrahedra masses.
Here we will give an overview of the multi-threaded mass deposit
approach and defer the CPU and GPU specific implementation 
details to the next Subsections.

In both cases, the worker threads keep running until all oct-tree leaf
nodes have been processed. The Lagrangian subblocks, see
Section~\ref{Section::DataInput}, that overlap
the spatial extent of the current leaf node, are distributed to the
available worker threads. We found that splitting them into 10 times 
more groups, than the number of worker threads, 
leads to good results in terms of dynamic load balancing between 
the worker threads. Each thread allocates its own private memory buffer
to accumulate the mass contributions. The mass deposit step involves
the computation of the exact intersection volumes between the
tetrahedra and the cubical grid cells and will be discussed in
detail in the Subsection~\ref{Sect-BoxTetIntersection}. 
Once a worker thread has deposited all mass associated with its tetrahedra, it is
assigned the next set of Lagrangian cell blocks and once all blocks
have been processed, the separate private mass buffers are added and
the worker threads signal the main thread that they are ready for new work.

\subsection{CPU Implementation}
\label{SubSec-CPU-Impl}
In the CPU case, each worker thread operates on one cell $L$ 
of its assigned Lagrangian grid blocks at a time. Its minimal,
axis-aligned bounding box is constructed and tested for intersection
with the extent of the leaf node's deposit grid patch. In the case of
no intersection, L's 6 tetrahedra are skipped and the next cell is
processed. Otherwise the first of the $6$ tetrahedra, 
$T_0$, is constructed, its minimal bounding box is determined 
and the worker thread loops over all cells $C_i$
of the deposit grid patch that overlap the bounding box of $T_0$.
For each cell $C_i$, we first test for trivial intersection configurations between  $C_i$
and $T_0$, i.~e.~no intersection, $C_i$ completely inside $T_0$ or
vice versa, which is implemented by computing the
relative orientation of the vertices of $T_0$ with respect to the axis-aligned faces of $C_i$.
Otherwise, the computationally more intensive algorithm discussed in
Subsection~\ref{Sect-BoxTetIntersection} is used to compute the exact
intersection volume. The resulting volume is transformed to its mass
equivalent, stored in the thread's private memory location for cell
$C_i$ and the next tetrahedron of $L$ is processed.

\subsection{GPU Implementation}
In the GPU case, we start one CPU thread for each available local GPU and
keep it running, until all leaf nodes of the grid structure have been
processed. Each CPU thread allocates private GPU memory to store the
Lagrangian grid block as well as a buffer to accumulate the mass contributions. 

On GPU achitectures, code divergence can severely degrade overall performance,
since sets of threads operate in lock-step and different
code paths causes subsets of the GPU threads to idle, resulting in an
underutilization of the available computational resources.
This poses a challenge for the geometry intersection, since the amount of work for the 
tetrahedra differs substantially, depending, for example, on how many
deposit cells are covered and on the number of non-trivial intersection
configurations.

On current {\sl NVIDIA} GPU architectures, sets of 32 threads, 
called {\sl lanes}, form a {\sl warp} and operate in lock-step. 
Hence, assigning one tetrahedron to each {\sl lane},
can result in only one of 32 threads performing useful
work. One potential solution would be to use different {\sl GPU kernels},
one that classifies the type of intersections
and buffers the results in global GPU memory, as well as two separate kernels that process the
trivial and non-trivial intersection cases, similar to the techniques
discussed for 2D triangle 
rasterization in~{\sl CUDA raster}~\cite{Laine:2011:HSR:2018323.2018337}.
However, in the 3-dimensional case, this would quickly exceed the available global GPU 
memory due to the drastically larger number of cells in the 3D deposit 
grids, as compared to the 2-dimensional framebuffers, as well as the
sheer size of the tesselations our approach targets.

\begin{figure}[h]
  \centering
  \includegraphics[width=.9\linewidth]{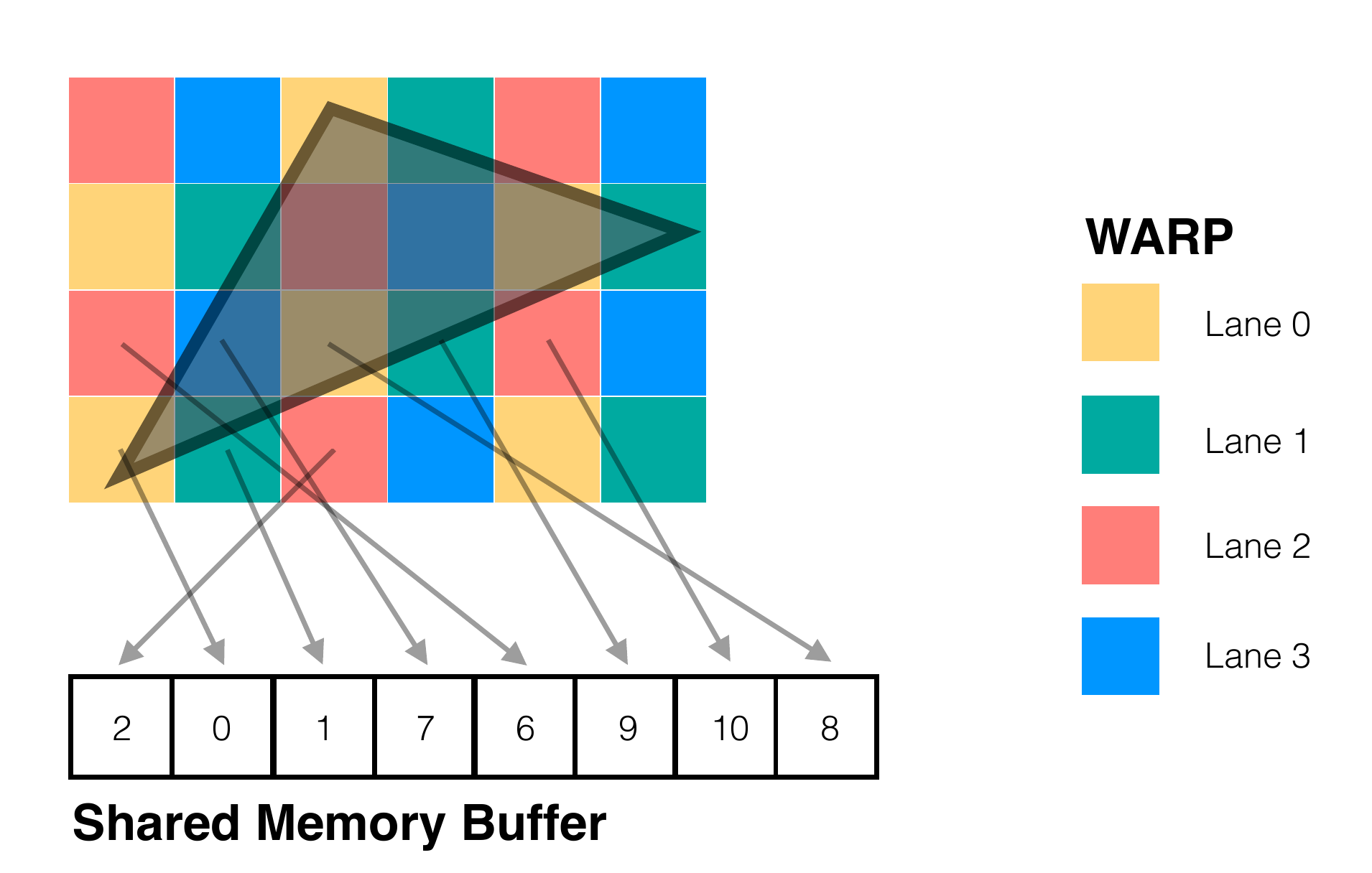}
  \caption{ 2D example of the GPU rasterization
    approach for a hypothetical warp size of 4. The lanes in the warp, depicted by
    different colors, classify 4 cells in lock-step. The IDs of
    cells intersecting the edges of the triangle are
    stored in a buffer in shared memory and processed in lock-step in
    a separate step. }
\label{Fig-GPURasta}
\end{figure}

We therefore follow a different approach, that avoids the usage of 
additional global memory by processing one tetrahedron per {\sl warp}, using a single GPU kernel.
Each warp picks a separate Lagrangian cell, tests its bounding box for intersection with the oct--tree node and, in case
of an intersection, constructs the first of the
$6$ tetrahedra. Each {\sl lane} picks a different cell of the
deposit grid patch and classifies the intersection configuration in
lock-step. Since the trivial cases are not computationally expensive and
cause only minimal thread divergence, they are processed immediately,
using an atomic operation to update the GPU's global memory
entry for this cell. However, in case a {\sl lane} detects
a non-trivial configuration, the more expensive intersection
routine discussed in Subsection~\ref{Sect-BoxTetIntersection} is not
called immediately. Instead the cell's ID is stored in the fast, low-latency {\sl shared
  memory}, as depicted using the 2D example in Figure~\ref{Fig-GPURasta}.
We use a buffer size of $64$ single precision integers, and determine
the current buffer position for each lane using an atomic counter in {\sl shared memory}. 
The lanes continue to classify cells, until the number of IDs, 
stored in the {\sl shared memory} buffer, exceeds
$31$. In this case, each {\sl lanes} picks one buffered ID and calls
the exact intersection routine for its cell. Once all lanes in the warp
returned from the call, they classify the next set of 32 cells in
lock-step. For our test cases, this approach reduced thread divergence substantially and 
performed about $20$ times faster compared to an alternative 
``one-tetrahedron per GPU thread'' approach. 

\subsection{Tetrahedron-Cube Intersection}     
\label{Sect-BoxTetIntersection}

\begin{figure*}
  \centering
  \includegraphics[width=0.24\textwidth]{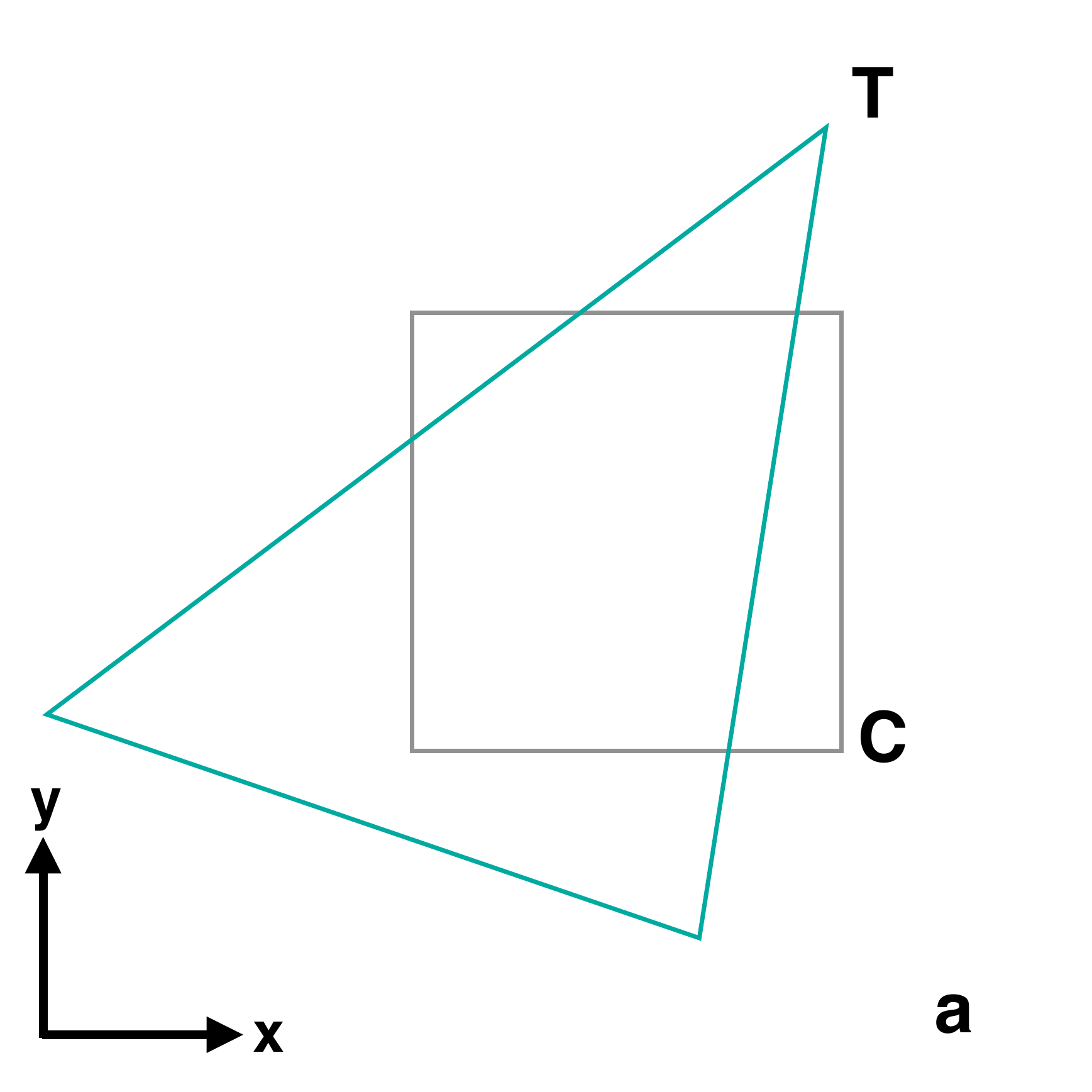}
\includegraphics[width=0.24\textwidth]{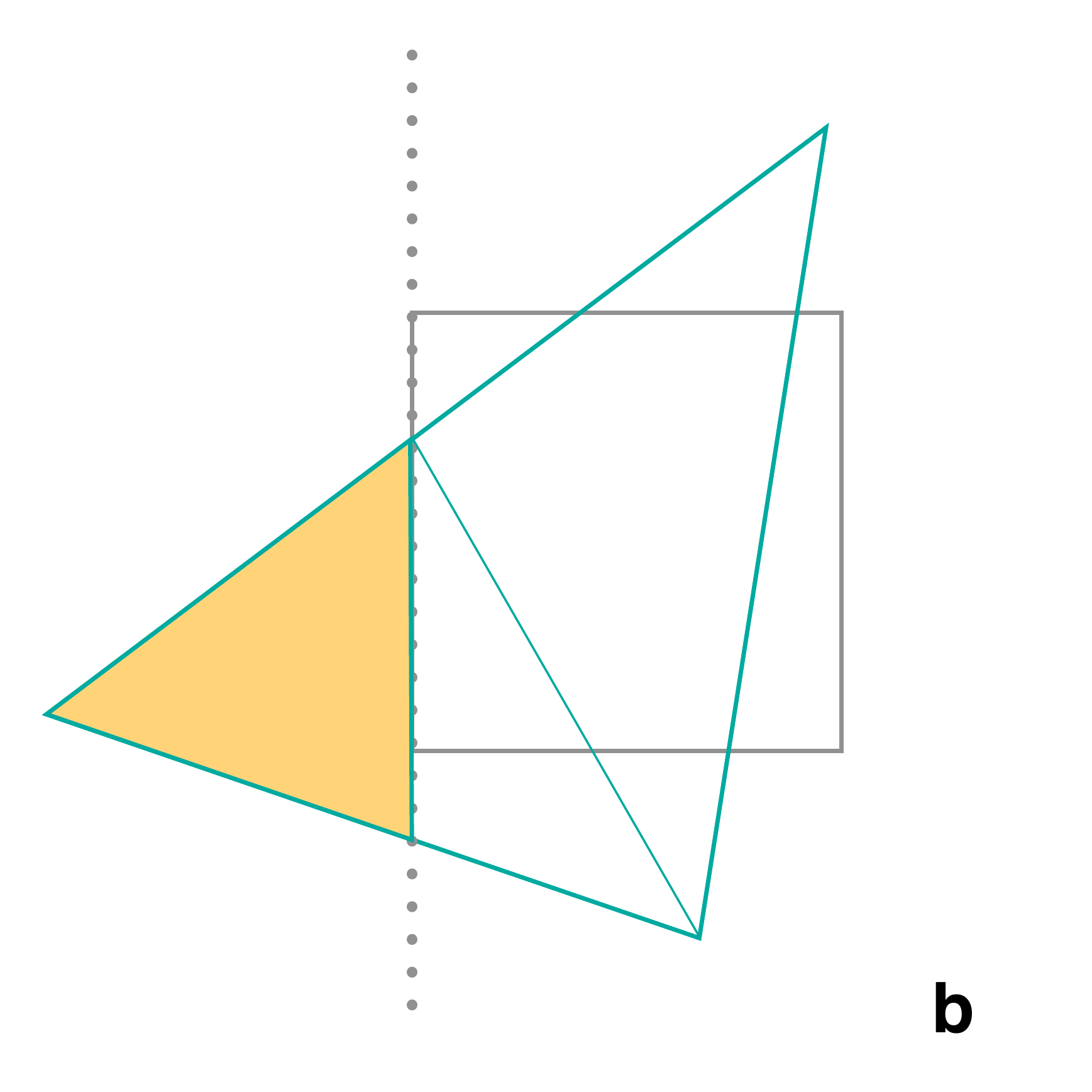}
\includegraphics[width=0.24\textwidth]{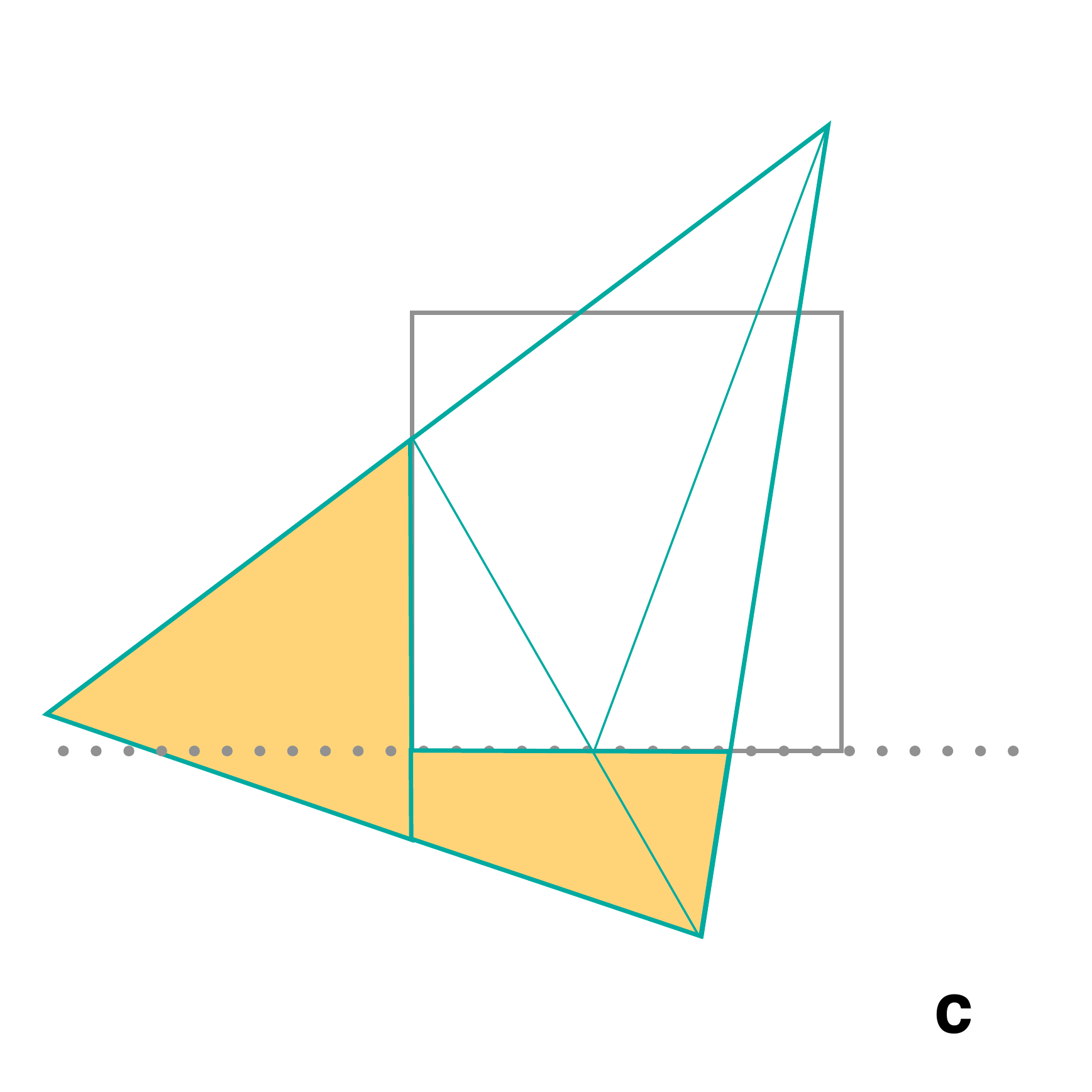}
\includegraphics[width=0.24\textwidth]{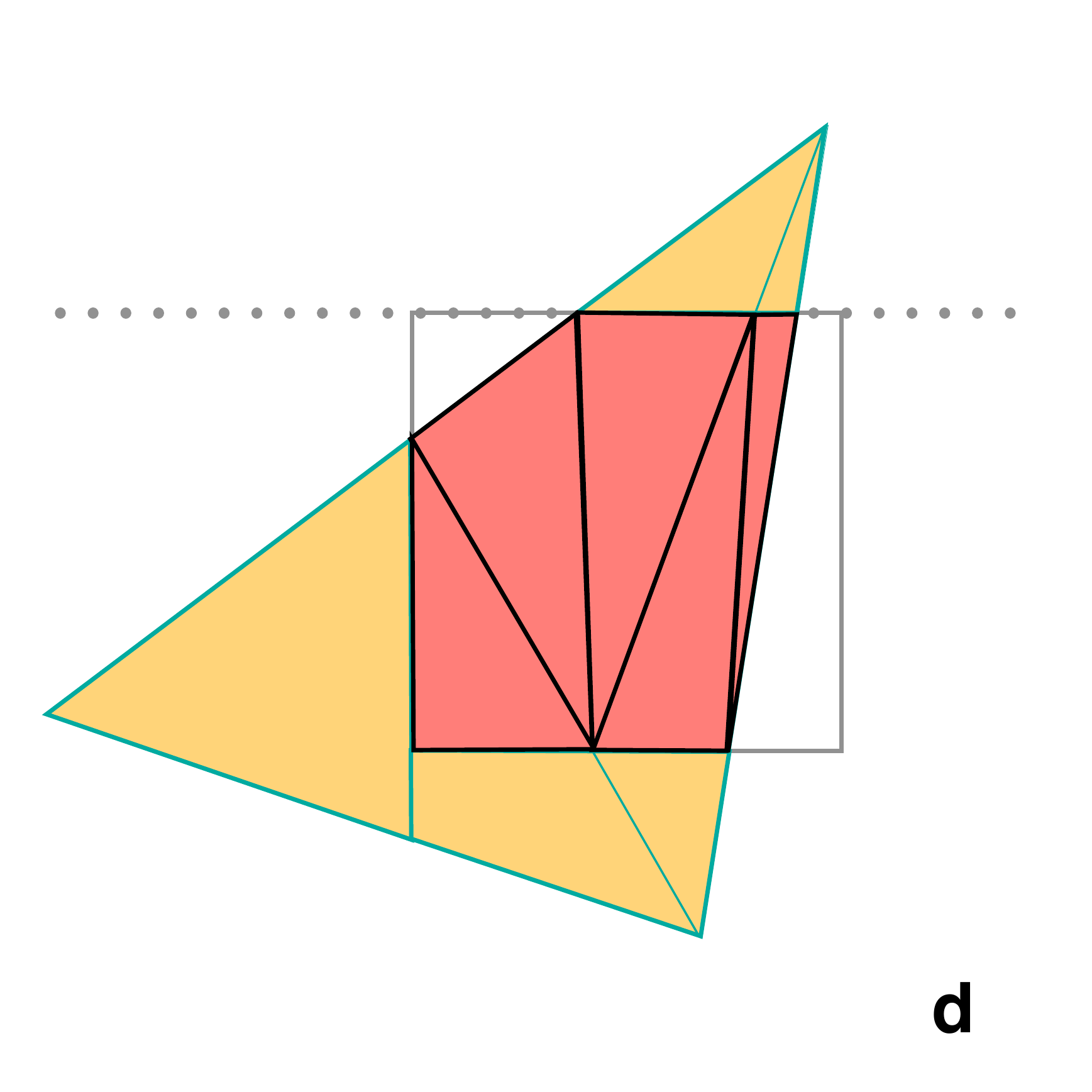}
\caption{ Computing the area of the intersection between cell $C$
and triangle $T$, by iteratively splitting $T$ against lines defined by
the edges of $C$, followed by a triangulation of the resulting polygon
on the half-space that includes $C$.}
\label{Fig-TetBoxIntersection}
\end{figure*}

In this Subsection, we describe our algorithm for computing the volume 
of the intersection between a tetrahedron $T$ and a cubical cell $C$.
For the sake of simplicity, we explain the algorithm in 2D, since the 
generalization to 3D is straightforward. 
So the task is to compute the area of the intersection between a triangle
$T$ and the quadratic cell $C$. Let us assume that the cell's lower
left vertex is at $(x=0,y=0)$ and its upper right at $(x=1,y=1)$, as
depicted in Figure~\ref{Fig-TetBoxIntersection}~(a).

The first step is to intersect $T$ with the line defined by
$(x=0,y)$. If the entire triangle is located on the half-plane
defined by $(x \leq 0, y)$, the intersection volume is zero and the
algorithm terminates. Otherwise the resulting polygon, located on the half-plane
$(x>0, y)$ that includes $C$, is triangulated by up to two triangles,
as shown in Figure~\ref{Fig-TetBoxIntersection}~(b). Next, the newly
generated triangles are tested for intersection with the line
$(x,y=0)$, the resulting interior region is triangulated and the
procedure is repeated for the lines $(x=1,y)$ and $(x,y=1)$. The
intersection area is computed by summing the areas of the resulting
triangles. The approach operates analogously in
$3$-dimensions, where the tetrahedron is iteratively split against the
2D planes defined by the faces of the cell.
In order to speed up the triangulation of auxiliary polygons, we precompute all
different cases and store them in a {\sl look-up table}. The location of
the vertices of $T$ relative to each plane defines a bitmask that serves as a
lookup index into the table, which returns the number of resulting
tetrahedra as well as the connectivity information to construct the
intermediate geometry.
Instead of carrying out the algorithm as sketched above for all three
dimensions, for the third dimension, we triangulate the parts of
the polygons outside $C$'s halfspace, instead the one inside, and subtract its 
volume from the interior tetrahedra, since this generally reduces 
the amount of generated geometry.

A possible GPU implementation is to cache the intermediate
tetrahedra and reuse them for each new dimension that is processed.
However, on the GPU, the fast {\sl shared memory} is too small to
store all geometry, so it would have to be moved to slower
global memory, substantially degrading overall performance.
Therefore, we avoid to cache intermediate tetrahedra and prune each tetrahedron
separately against all cell's faces, before the next intermediate
tetrahedron is processed.  Although this involves redundant
computations, it turns out to increase performance on
massively parallel graphics hardware, where data movement is
relatively expensive compared to computation.

\section{Results and Discussions}
\label{Sec-Results}

We conducted our experiments on the  {\sl Sherlock} and {\sl XStream}
HPC~clusters at the Stanford Research Computing Center~(SRCC).
{\sl Sherlock's} hardware resources include 120 general compute nodes,
each with a dual socket {\sl Intel Xeon CPU E5-2650 v2} at $2.60$~GHz, 
8~cores per socket and $64$ GB DDR3 RAM. 
{\sl XStream} consists of 65 compute nodes, each equipped with 2~x~Intel~Xeon~CPU
E5-2680~v2 @ 2.80GHz (10 cores per socket), 256~GB of RAM and 8~x~NVIDIA~Tesla~K80 (16
logical GPUs). Both clusters are interconnected with FDR~Infiniband
and have access to a PB sized Lustre parallel file system.
For our measurements we used simulation snapshots  with  $2048^3$ and
$4096^3$ particles at redshift $0$ from the {\sl Dark Sky Simulations: Early Data Release}~\cite{Skillman:2014qca}.

\subsection{Strong Scaling}
\begin{figure}[h]

\begin{center}
 \centering
  \includegraphics[width=1.\linewidth]{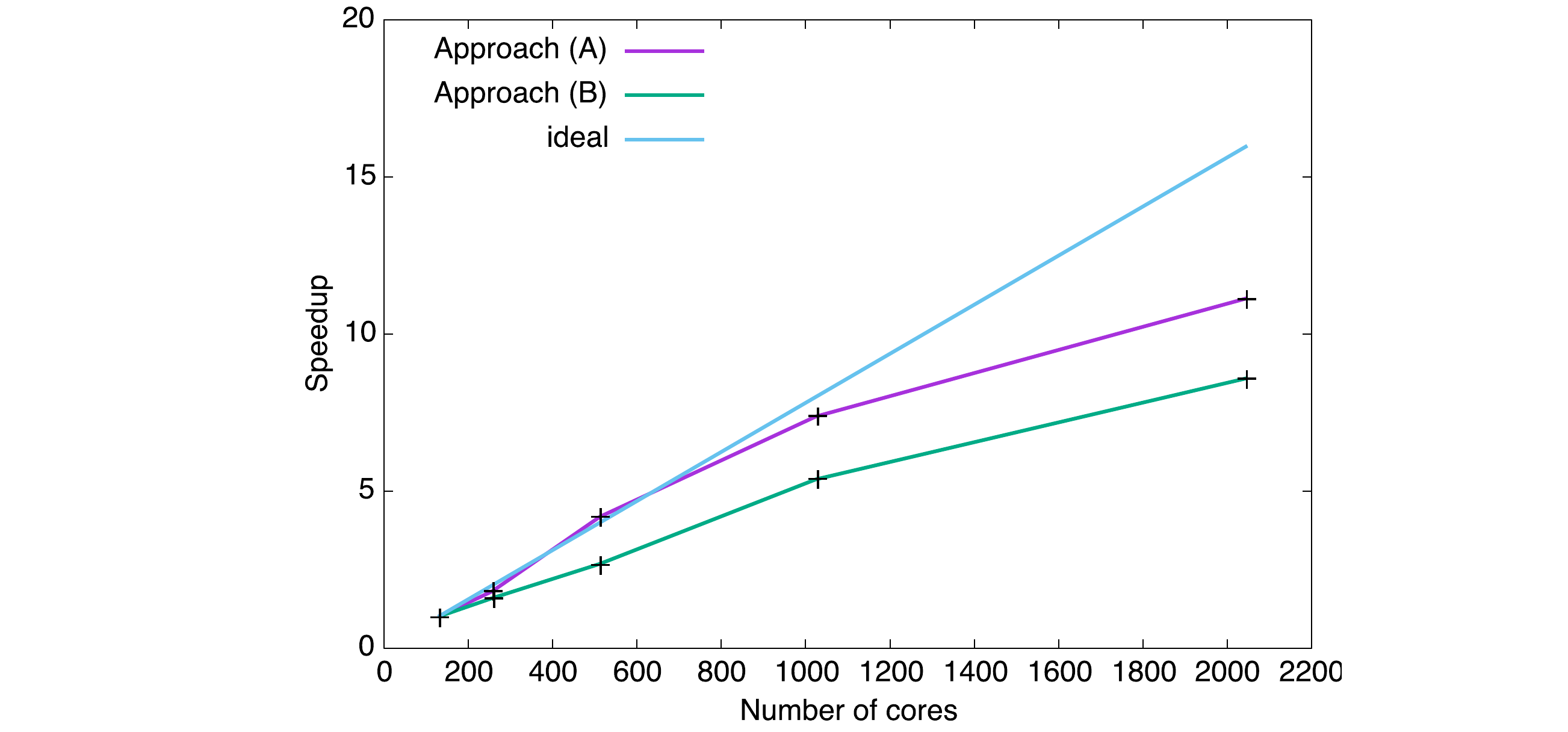}
  \caption{Strong scaling results for $2048^3$ particles.}
  \label{Fig::Strong_Scaling}
\end{center}
\end{figure}

The strong scaling tests utilized between $128$ and $2048$ CPU
cores on {\sl Sherlock}, with one MPI rank per compute node and
$8$ threads per MPI rank. The test problem was the exact mass deposit
of a $2048^3$ particles \nb simulation onto a fully refined octree with 4 levels of
refinement and $128^3$ cells per node, resulting in a total grid
resolution of $2048^3$ cells. Tables~\ref{Tabl::StrongScalingComp1}
and \ref{Tabl::StrongScalingComp2} as well as
Figure~\ref{Fig::Strong_Scaling} summarize the results for the two different
communication strategies discussed in Section~\ref{Sec-LoadBalancing} - exchanging point
data~A versus exchanging deposit grid arrays between cluster
nodes~B.~\footnote{The times for the 
initial data I/O  and distributed tessellation construction
(Section~\ref{Section::DataInput}), as well as the
octree structure (Section~\ref{Section::AdaptiveGridStructure})
were small compared to the overall runtime (in the order of 1\%), so we did not
list them separately here.}.

For approach~B, up to $5000$ leaf node mass arrays per compute node
were locally cached on each process.
As shown in tables~\ref{Tabl::StrongScalingComp1} and
\ref{Tabl::StrongScalingComp2},  the performance of
approach~A was between $1.2$ and $1.9$ times faster. 
Also in terms of scalability, approach~A, which shows almost
perfect scaling up to $1024$ cores, outperformed~B. For
$2048$ cores, the speed-up was reduced from a theoretical factor of 
16 to about 11, due to increased overhead of communication between 
the cluster nodes.

\begin{figure}[h]
\small
\begin{center}
 \begin{tabular}{ | c | c | c | }
  \hline	
\# CPU cores & Time (min) & Speedup\\
\hline \hline		
  128   & 603 & 1.0 \\
  256   & 331 & 1.8 \\
  512   & 143 & 4.2 \\
  1024 & 81   & 7.4 \\
  2048 & 53   & 11.2 \\
  \hline  
\end{tabular}
\caption{Strong scaling results for communication strategy~A for depositing
the mass associated with a $2048^3$ particles \nb simulation onto
a fully refined octree with 4 levels of refinement and $128^3$ cells
per node, resulting in a total grid resolution of $2048^3$ cells. }
\label{Tabl::StrongScalingComp1}
\end{center}
\end{figure}

\begin{figure}[h]
\small
\begin{center}
 \begin{tabular}{ | c | c | c | }
  \hline	
\# cores & Time (min) & Speedup\\
\hline \hline		
  128   & 712 & 1.0 \\
  256   & 458 & 1.6 \\
  512   & 266 & 2.7 \\
  1024 & 129 & 5.0  \\
  2048 & 77   & 9.2 \\
  \hline  
\end{tabular}
\caption{Strong scaling results for communication alternative~B, using the
  same test problem as in Table~\ref{Tabl::StrongScalingComp1}. }
\label{Tabl::StrongScalingComp2}
\end{center}
\end{figure}

\subsection{Weak Scaling}
\begin{figure}[h]
  \centering
  \includegraphics[width=1.\linewidth]{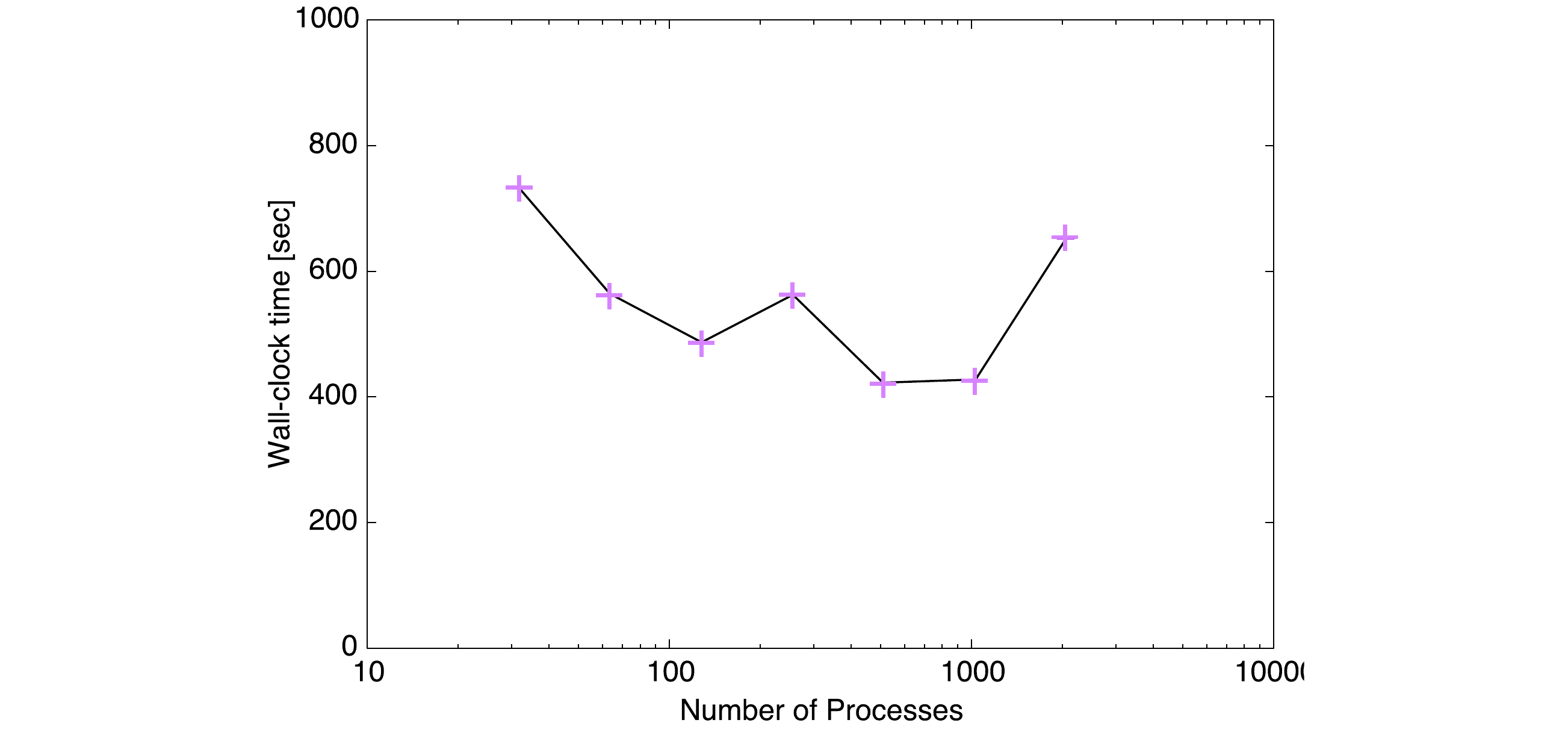}
 \caption{
    Weak scaling results for a mass deposit problem using $512^3$ to $2048^3$
    particles, using between $32$ and $2048$ processes.
}
\label{Fig::WeakScaling}
\end{figure}

In order to ensure that the complexity of the workload for the different
datasets used for the weak scaling test scales linearly with the
number of utilized cores, the different datasets were generated
analytically, by placing the dark matter particles onto a uniform grid
and adding randomized positive or negative offsets of up to one cell
width in each direction to the particle positions. The number of particles and 
the number of cells in the deposit grid was successively increased by a factor of 
two, ranging from $512^3$ to $2048^3$ particles, with $256^3$ to
$1024^3$ cells in the deposit grid structure. The datasets were
processed utilizing between $32$ and $2048$ cores on
$1$ up to $64$ nodes on the Sherlock cluster, doubling the number 
of nodes between consecutive tasks. The resulting 
wall clock times for communication strategy~(A) are 
given in Figure~\ref{Fig::WeakScaling}.
 
\subsection{GPU Performance}
\begin{figure}[h]
  \centering
  \includegraphics[width=1.\linewidth]{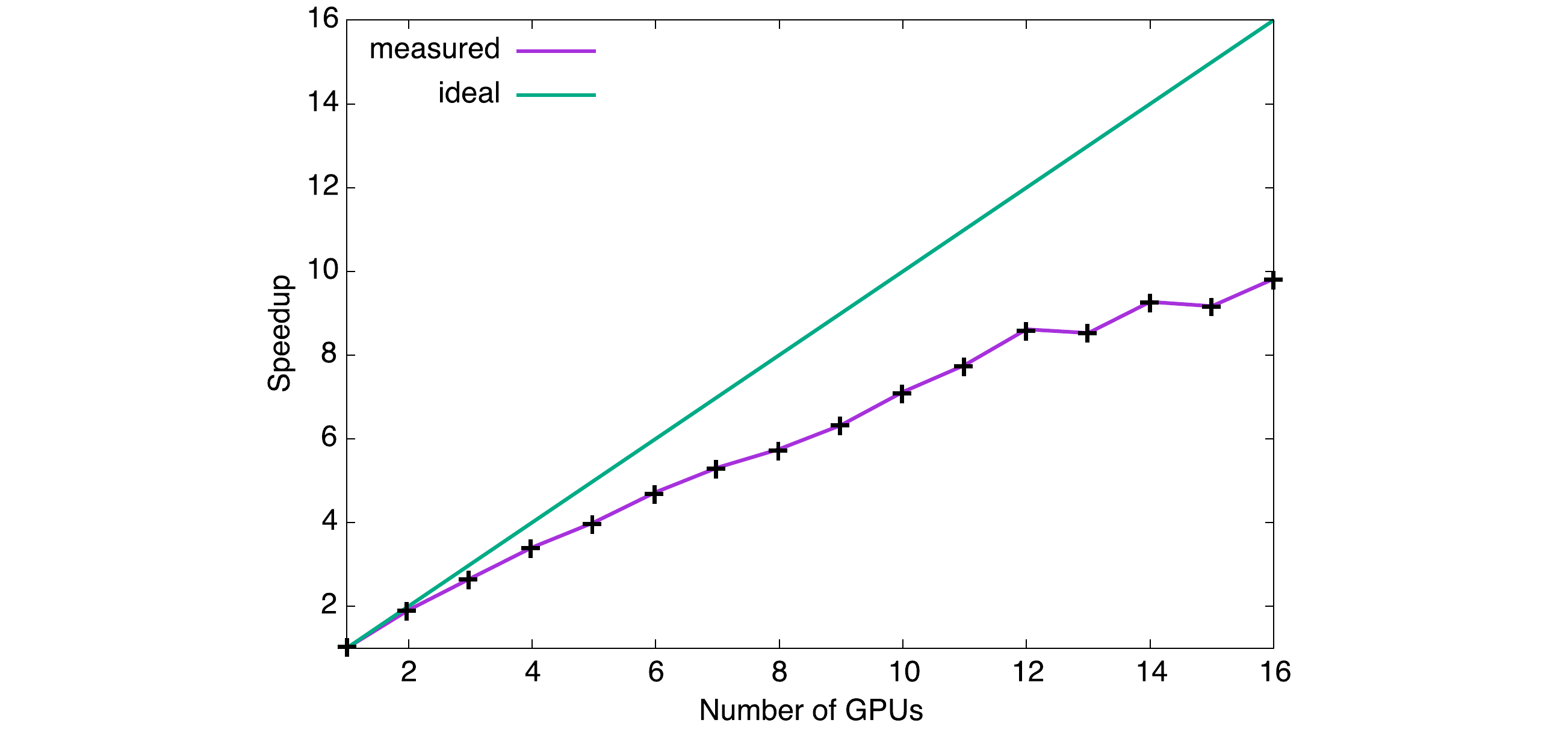}
\caption{
   Ideal and measured GPU strong scaling results for $1024^3$ particles and
   $1$ to $16$ GPUs.
}
\label{Fig::GPUScaling}
\end{figure}

The performance between the CPU and GPU version
of the code discussed in Section~\ref{Sec-MassDeposit}, was conducted on one node of the {\sl XStream}
GPU cluster, with two Intel~Xeon~Processor E5-2680 v2 CPUs and
one {\sl Nvidia K80} card with two Kepler~GK210~GPUs. The peak performance for the CPU version was achieved with $40$
threads ($2 \time 10$ cores with hyperthreading). The test problem was the mass deposit for a
\nb simulation with $2048^3$ particles onto a fully refined octree with 4 levels of
refinement and $128^3$ cells per node, resulting in a total grid
resolution of $2048^3$ cells. The CPU version took about $1995$
minutes to complete, whereas the GPU version was finished in about
$230$ minutes, so the achieved GPU speedup was about $8.7$.
Figure~\ref{Fig::GPUScaling} and Table~\ref{Tabl::GPUScaling} show the result of a strong scaling test on a
single node of the {\sl XStream} cluster, utilizing between $1$ and all $16$
available GPUs. The test dataset was a $1024^3$ particle
dataset. The speedup increased roughly linearly with the number of
GPUs with a slope of about $0.63$, due to the increased overhead of
transferring the point data from the CPU to several GPUs and reducing
the local deposit grid arrays back on the CPU.

\begin{figure}[h]
\small
\begin{center}
 \begin{tabular}{ | c | c | c | }
  \hline	
\# GPUs & Time (sec) & Speedup\\
\hline \hline
1    &   4144 & 1.0 \\ 
2    &   2170 & 1.9 \\
3    &   1556 & 2.7 \\
4    &   1217 & 3.4 \\
5    &   1035 & 4.0 \\
6    &   877 & 4.7 \\
7    &   780 & 5.3 \\
8    &   720 & 5.8 \\
9    &   655 & 6.3 \\
10   &   582 & 7.1 \\
11   & 534 & 7.8 \\
12   &   481 & 8.6 \\
13   &   486 & 8.5 \\
14   &   447 & 9.3 \\
15   &   452 & 9.2 \\
16   &   422 & 9.8 \\
  \hline  
\end{tabular}
\caption{Strong scaling results for up to 16 GPUs. }
\label{Tabl::GPUScaling}
\end{center}
\end{figure}

\begin{figure*}[htb]
  \centering
  \includegraphics[width=0.7\textwidth]{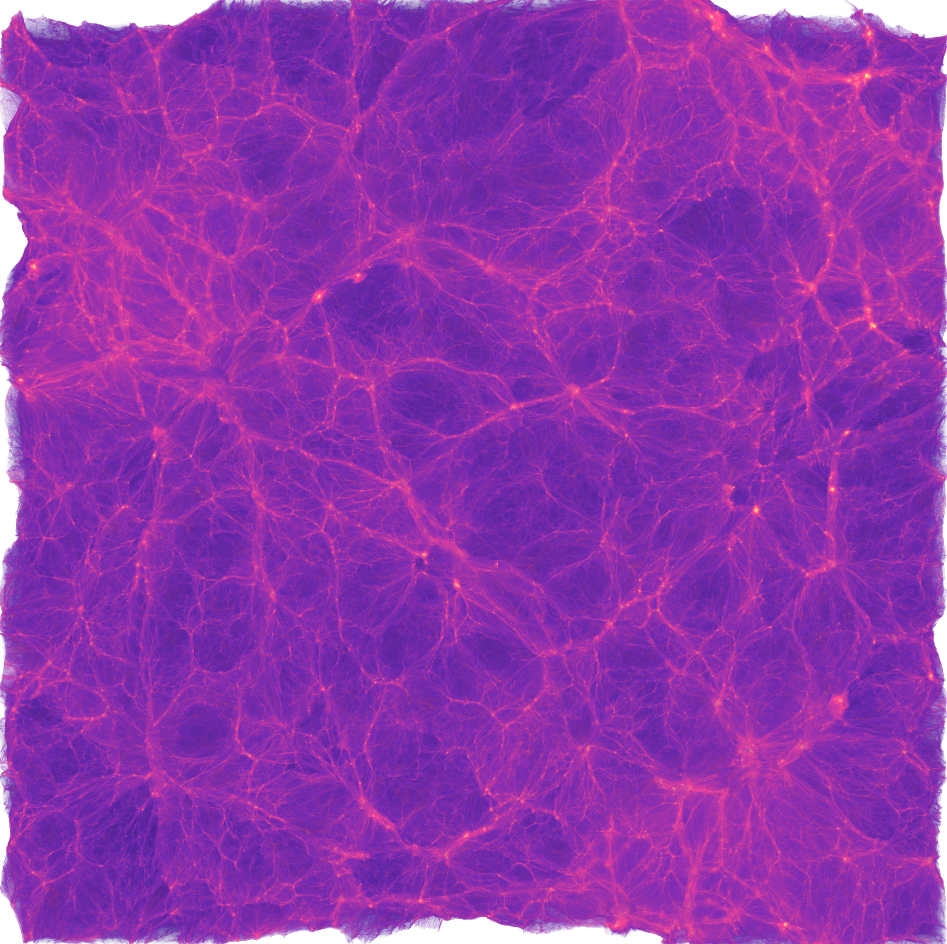}
\includegraphics[width=0.7\textwidth]{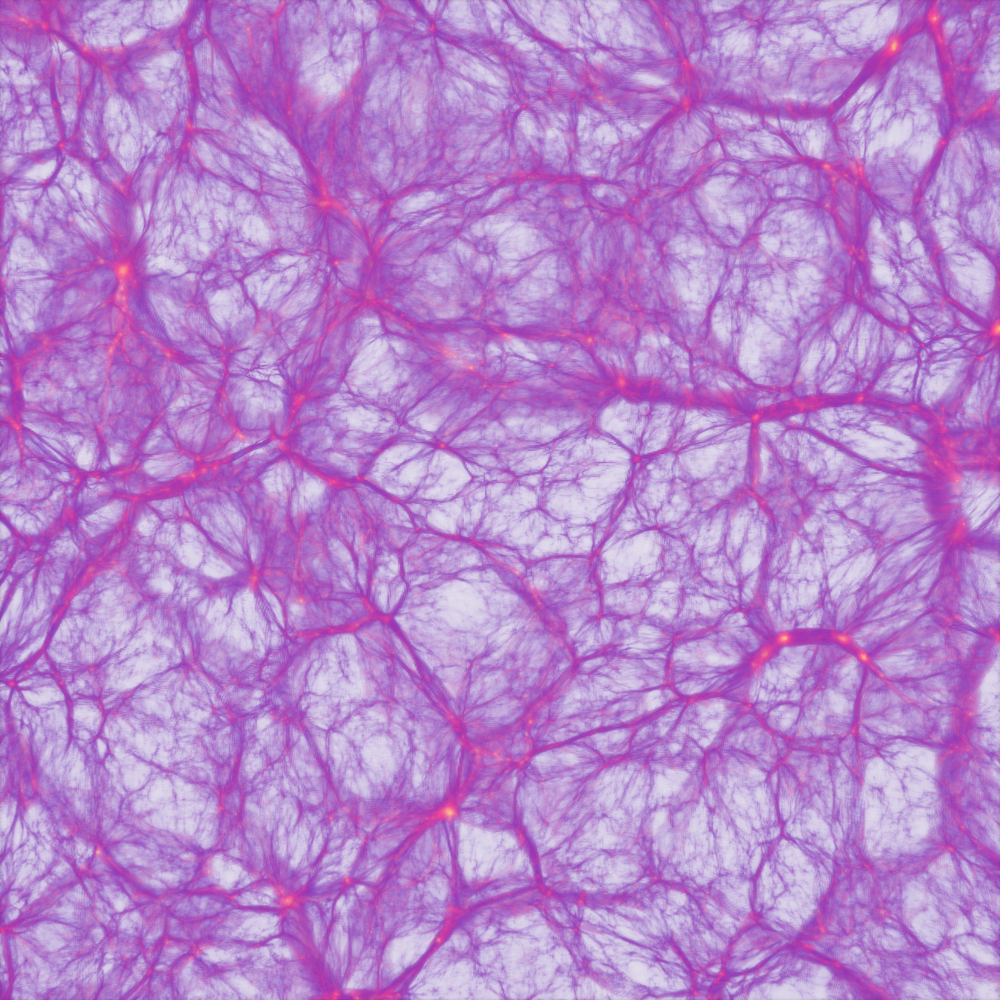}
\caption{ Direct volume rendering of a \nb
  simulation with $4096^3$ particles. The mass was resampled onto an
  oct-tree data structure with an effective resolution of $8192^3$
  cells. The upper part shows the whole domain, whereas the lower 
  one depicts a zoomed view of a thin slice through the volume. }
\label{Fig-VolumeRendering}
\end{figure*}

We further performed an adaptive mass deposit test using a $4096^3$
particle \nb dataset, using a leaf node size of $64^3$ cells per
node and a maximum of $7$~refinement levels, resulting in an effective resolution
of $8192^3$ cells on the highest level of resolution.
The test was run on 16 compute nodes of {\sl XStream}, utilized all available 8
{\sl Nvidia K80} cards per node and finished in $46$ minutes. 
Figure~\ref{Fig-VolumeRendering} shows a visualization of the
multi-resolution density field with direct volume rendering.

\section{Summary and Future Work}
\label{Sec:Discussion}
We presented an approach for the distributed computation of accurate density
fields from large \nb dark matter simulations, using the phase-space
element technique, that is tailored to massively parallel cluster architectures
equipped with GPU accelerators. We employed an oct-tree grid structure to
sample the densities onto cubical cells, adapting the resolution
according to the features of the underlying tessellation.
We introduced a GPU algorithm for the computationally intensive
step of geometrically intersecting the tetrahedra with the cubical
cells of the octree, which achieves almost an order of magnitude speedup
compared to an optimized CPU implementation.
We further presented two dynamic load balancing
strategies and demonstrated the overall good weak and strong scaling
performance of our approach with several large datasets. 
Our measurements showed that it is beneficial to communicate particle data,
instead of reducing deposit grid patches in terms of overall performance
and scaling features.

We see various directions for future work.
It would be interesting to compare the performance
of the GPU implementation with one optimized for {\sl Intel Xeon Phi} coprocessors. 
We would also like to investigate if the proposed solution can be extended
and applied to {\sl in-situ} scenarios, i.e., performed alongside the
underlying simulation, by sharing the available hardware resources.
Furthermore, in addition to the mass deposit onto the leafs nodes of the oct-tree structure,
generating coarser representations for the internal nodes  
would be beneficial for interactive visualization approaches,
like direct volume rendering, that employ multi-resolution
representations of the data to achieve interactive frame rates.
Our implementation is open source and available at {\bf https://github.com/kaehlerr/ADECO}.

\section{Acknowledgments}
We are very grateful to Tom Abel and Devon Powell for useful discussions on
this topic. We would also like to thank the {\sl Dark Sky Simulations} collaboration for
making their data publically available, in particular Sam Skillman for his advice
regarding the data format.  This work used the {\sl Sherlock} and {\sl XStream}
clusters, hosted at the Stanford Research Computing Center, which is supported by the 
National Science Foundation Major Research Instrumentation program
(ACI-1429830). We would also like to thank Nvidia for supporting KIPAC's
computing department with several GPUs.

\begin{figure*}[htb]
  \centering
  \includegraphics[width=0.95\textwidth]{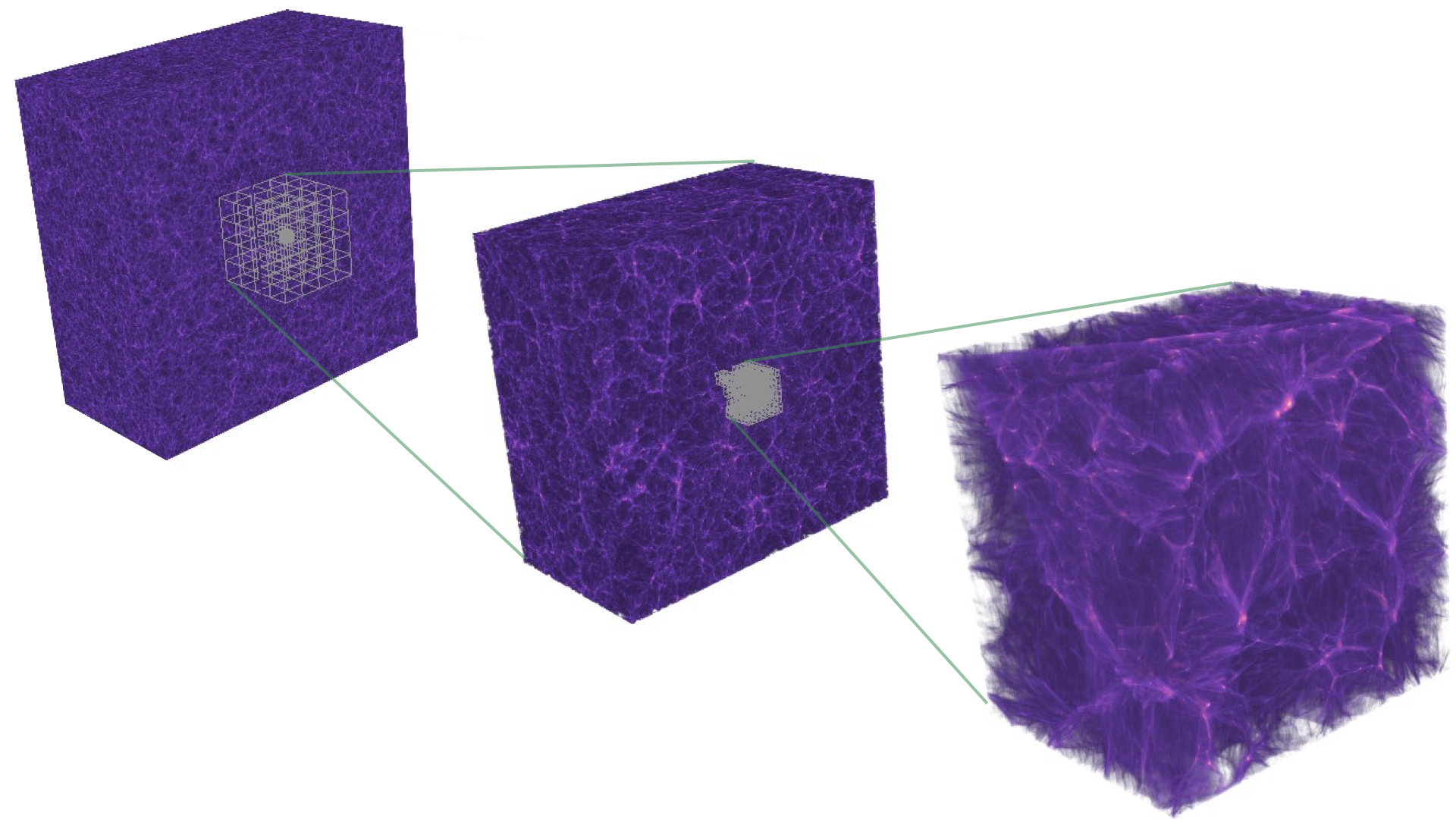}
\caption{ Direct volume rendering of a \nb
  simulation with $4096^3$ particles, zooming from a region of $1600$ Mpc
  down to a region of about $6$ Mpc. The mass was resampled onto an
  oct-tree data structure with $9$ levels of resolution. }
\label{Fig-VolumeRenderingZoom}
\end{figure*}

%% file: ASCOM_2016_87_revision.bbl
\begin{thebibliography}{10}
\expandafter\ifx\csname url\endcsname\relax
  \def\url#1{\texttt{#1}}\fi
\expandafter\ifx\csname urlprefix\endcsname\relax\def\urlprefix{URL }\fi
\expandafter\ifx\csname href\endcsname\relax
  \def\href#1#2{#2} \def\path#1{#1}\fi

\bibitem{Dikaiakos:1996:PSC:237578.237590}
M.~D. Dikaiakos, J.~Stadel, \href{http://doi.acm.org/10.1145/237578.237590}{A
  performance study of cosmological simulations on message-passing and
  shared-memory multiprocessors}, in: Proceedings of the 10th International
  Conference on Supercomputing, ICS '96, ACM, New York, NY, USA, 1996, pp.
  94--101.
\newblock \href {http://dx.doi.org/10.1145/237578.237590}
  {\path{doi:10.1145/237578.237590}}.
\newline\urlprefix\url{http://doi.acm.org/10.1145/237578.237590}

\bibitem{0067-0049-111-1-73}
A.~V. Kravtsov, A.~A. Klypin, A.~M. Khokhlov,
  \href{http://stacks.iop.org/0067-0049/111/i=1/a=73}{Adaptive refinement tree:
  A new high-resolution n-body code for cosmological simulations}, The
  Astrophysical Journal Supplement Series 111~(1) (1997) 73.
\newline\urlprefix\url{http://stacks.iop.org/0067-0049/111/i=1/a=73}

\bibitem{Teyssier:2001cp}
R.~Teyssier, {Cosmological hydrodynamics with adaptive mesh refinement: a new
  high resolution code called ramses}, Astron. Astrophys. 385 (2002) 337--364.
\newblock \href {http://arxiv.org/abs/astro-ph/0111367}
  {\path{arXiv:astro-ph/0111367}}, \href
  {http://dx.doi.org/10.1051/0004-6361:20011817}
  {\path{doi:10.1051/0004-6361:20011817}}.

\bibitem{Dubinski:2004}
J.~Dubinski, J.~Kimb, C.~Park, R.~Humbleb, {GOTPM: a parallel hybrid
  particle-mesh treecode}, New Astronomy 9 (2004) 111--126.
\newblock \href {http://dx.doi.org/10.1016/j.newast.2003.08.002}
  {\path{doi:10.1016/j.newast.2003.08.002}}.

\bibitem{Springel05thecosmological}
V.~Springel, {The Cosmological Simulation Code Gadget-2}, Monthly Notices of
  the Royal Astronomical Society 364.

\bibitem{0067-0049-184-2-298}
M.~Wetzstein, A.~F. Nelson, T.~Naab, A.~Burkert,
  \href{http://stacks.iop.org/0067-0049/184/i=2/a=298}{Vine¿a numerical code
  for simulating astrophysical systems using particles. i. description of the
  physics and the numerical methods}, The Astrophysical Journal Supplement
  Series 184~(2) (2009) 298.
\newline\urlprefix\url{http://stacks.iop.org/0067-0049/184/i=2/a=298}

\bibitem{doi:10.1093/pasj/61.6.1319}
T.~Ishiyama, T.~Fukushige, J.~Makino, \href{+
  http://dx.doi.org/10.1093/pasj/61.6.1319}{Greem: Massively parallel treepm
  code for large cosmological n-body simulations}, Publications of the
  Astronomical Society of Japan 61~(6) (2009) 1319.
\newblock \href
  {http://arxiv.org/abs//oup/backfile/content_public/journal/pasj/61/6/10.1093/pasj/61.6.1319/2/pasj61-1319.pdf}
  {\path{arXiv:/oup/backfile/content_public/journal/pasj/61/6/10.1093/pasj/61.6.1319/2/pasj61-1319.pdf}},
  \href {http://dx.doi.org/10.1093/pasj/61.6.1319}
  {\path{doi:10.1093/pasj/61.6.1319}}.
\newline\urlprefix\url{+ http://dx.doi.org/10.1093/pasj/61.6.1319}

\bibitem{Habib:2013:HES:2503210.2504566}
S.~Habib, V.~Morozov, N.~Frontiere, H.~Finkel, A.~Pope, K.~Heitmann,
  \href{http://doi.acm.org/10.1145/2503210.2504566}{Hacc: Extreme scaling and
  performance across diverse architectures}, in: Proceedings of the
  International Conference on High Performance Computing, Networking, Storage
  and Analysis, SC '13, ACM, New York, NY, USA, 2013, pp. 6:1--6:10.
\newblock \href {http://dx.doi.org/10.1145/2503210.2504566}
  {\path{doi:10.1145/2503210.2504566}}.
\newline\urlprefix\url{http://doi.acm.org/10.1145/2503210.2504566}

\bibitem{Warren:2013:IPH:2503210.2503220}
M.~S. Warren, \href{http://doi.acm.org/10.1145/2503210.2503220}{2hot: An
  improved parallel hashed oct-tree n-body algorithm for cosmological
  simulation}, in: Proceedings of the International Conference on High
  Performance Computing, Networking, Storage and Analysis, SC '13, ACM, New
  York, NY, USA, 2013, pp. 72:1--72:12.
\newblock \href {http://dx.doi.org/10.1145/2503210.2503220}
  {\path{doi:10.1145/2503210.2503220}}.
\newline\urlprefix\url{http://doi.acm.org/10.1145/2503210.2503220}

\bibitem{2014ApJS..211...19B}
G.~L. {Bryan}, M.~L. {Norman}, B.~W. {O'Shea}, T.~{Abel}, J.~H. {Wise}, M.~J.
  {Turk}, D.~R. {Reynolds}, D.~C. {Collins}, P.~{Wang}, S.~W. {Skillman},
  B.~{Smith}, R.~P. {Harkness}, J.~{Bordner}, J.-h. {Kim}, M.~{Kuhlen},
  H.~{Xu}, N.~{Goldbaum}, C.~{Hummels}, A.~G. {Kritsuk}, E.~{Tasker},
  S.~{Skory}, C.~M. {Simpson}, O.~{Hahn}, J.~S. {Oishi}, G.~C. {So}, F.~{Zhao},
  R.~{Cen}, Y.~{Li}, {Enzo Collaboration}, {ENZO: An Adaptive Mesh Refinement
  Code for Astrophysics}, apjs 211 (2014) 19.
\newblock \href {http://arxiv.org/abs/1307.2265} {\path{arXiv:1307.2265}},
  \href {http://dx.doi.org/10.1088/0067-0049/211/2/19}
  {\path{doi:10.1088/0067-0049/211/2/19}}.

\bibitem{0004-637X-765-1-39}
A.~S. Almgren, J.~B. Bell, M.~J. Lijewski, Z.~Luki¿, E.~V. Andel,
  \href{http://stacks.iop.org/0004-637X/765/i=1/a=39}{Nyx: A massively parallel
  amr code for computational cosmology}, The Astrophysical Journal 765~(1)
  (2013) 39.
\newline\urlprefix\url{http://stacks.iop.org/0004-637X/765/i=1/a=39}

\bibitem{Hockney:1988:CSU:62815}
R.~W. Hockney, J.~W. Eastwood, Computer Simulation Using Particles, Taylor \&
  Francis, Inc., Bristol, PA, USA, 1988.

\bibitem{1969JCoPh...3..494B}
C.~K. {Birdsall}, D.~{Fuss}, {Clouds-in-clouds, clouds-in-cells physics for
  many-body plasma simulation}, Journal of Computational Physics 3 (1969)
  494--511.
\newblock \href {http://dx.doi.org/10.1016/0021-9991(69)90058-8}
  {\path{doi:10.1016/0021-9991(69)90058-8}}.

\bibitem{1988CoPhC..48...89M}
J.~J. {Monaghan}, {An introduction to SPH}, Computer Physics Communications 48
  (1988) 89--96.
\newblock \href {http://dx.doi.org/10.1016/0010-4655(88)90026-4}
  {\path{doi:10.1016/0010-4655(88)90026-4}}.

\bibitem{2008MNRAS.386.2101N}
M.~C. {Neyrinck}, {ZOBOV: a parameter-free void-finding algorithm}, mnras 386
  (2008) 2101--2109.
\newblock \href {http://arxiv.org/abs/0712.3049} {\path{arXiv:0712.3049}},
  \href {http://dx.doi.org/10.1111/j.1365-2966.2008.13180.x}
  {\path{doi:10.1111/j.1365-2966.2008.13180.x}}.

\bibitem{Shandarin:2011jv}
S.~Shandarin, S.~Habib, K.~Heitmann,
  \href{http://link.aps.org/doi/10.1103/PhysRevD.85.083005}{Cosmic web,
  multistream flows, and tessellations}, Phys. Rev. D 85 (2012) 083005.
\newblock \href {http://dx.doi.org/10.1103/PhysRevD.85.083005}
  {\path{doi:10.1103/PhysRevD.85.083005}}.
\newline\urlprefix\url{http://link.aps.org/doi/10.1103/PhysRevD.85.083005}

\bibitem{2011AbelHahnKaehler}
T.~Abel, O.~Hahn, R.~Kaehler,
  \href{http://mnras.oxfordjournals.org/content/427/1/61.abstract}{Tracing the
  dark matter sheet in phase space}, Monthly Notices of the Royal Astronomical
  Society 427~(1) (2012) 61--76.
\newblock \href
  {http://arxiv.org/abs/http://mnras.oxfordjournals.org/content/427/1/61.full.pdf+html}
  {\path{arXiv:http://mnras.oxfordjournals.org/content/427/1/61.full.pdf+html}},
  \href {http://dx.doi.org/10.1111/j.1365-2966.2012.21754.x}
  {\path{doi:10.1111/j.1365-2966.2012.21754.x}}.
\newline\urlprefix\url{http://mnras.oxfordjournals.org/content/427/1/61.abstract}

\bibitem{Hahn:2012ma}
O.~Hahn, T.~Abel, R.~Kaehler, {A new approach to simulating collisionless dark
  matter fluids}, Mon. Not. Roy. Astron. Soc. 434 (2013) 1171.
\newblock \href {http://arxiv.org/abs/1210.6652} {\path{arXiv:1210.6652}},
  \href {http://dx.doi.org/10.1093/mnras/stt1061}
  {\path{doi:10.1093/mnras/stt1061}}.

\bibitem{10.1109..TVCG.2012.187}
R.~Kaehler, O.~Hahn, T.~Abel, A novel approach to visualizing dark matter
  simulations, IEEE Transactions on Visualization and Computer Graphics 18~(12)
  (2012) 2078--2087.
\newblock \href
  {http://dx.doi.org/doi.ieeecomputersociety.org/10.1109/TVCG.2012.187}
  {\path{doi:doi.ieeecomputersociety.org/10.1109/TVCG.2012.187}}.

\bibitem{Igouchkine:2016:VRD:3002151.3002163}
O.~Igouchkine, N.~Leaf, K.-L. Ma,
  \href{http://doi.acm.org/10.1145/3002151.3002163}{Volume rendering dark
  matter simulations using cell projection and order-independent transparency},
  in: SIGGRAPH ASIA 2016 Symposium on Visualization, SA '16, ACM, New York, NY,
  USA, 2016, pp. 8:1--8:8.
\newblock \href {http://dx.doi.org/10.1145/3002151.3002163}
  {\path{doi:10.1145/3002151.3002163}}.
\newline\urlprefix\url{http://doi.acm.org/10.1145/3002151.3002163}

\bibitem{Angulo:2013bfq}
R.~E. Angulo, R.~Chen, S.~Hilbert, T.~Abel, {Towards noiseless gravitational
  lensing simulations}, Mon. Not. Roy. Astron. Soc. 444~(3) (2014) 2925--2937.
\newblock \href {http://arxiv.org/abs/1309.1161} {\path{arXiv:1309.1161}},
  \href {http://dx.doi.org/10.1093/mnras/stu1608}
  {\path{doi:10.1093/mnras/stu1608}}.

\bibitem{Hahn:2014lca}
O.~Hahn, R.~E. Angulo, T.~Abel, {The Properties of Cosmic Velocity Fields},
  Mon. Not. Roy. Astron. Soc. 454~(4) (2015) 3920--3937.
\newblock \href {http://arxiv.org/abs/1404.2280} {\path{arXiv:1404.2280}},
  \href {http://dx.doi.org/10.1093/mnras/stv2179}
  {\path{doi:10.1093/mnras/stv2179}}.

\bibitem{Hahn:2015sia}
O.~Hahn, R.~E. Angulo, {An adaptively refined phase space element method for
  cosmological simulations and collisionless dynamics}, Mon. Not. Roy. Astron.
  Soc. 455~(1) (2016) 1115--1133.
\newblock \href {http://arxiv.org/abs/1501.01959} {\path{arXiv:1501.01959}},
  \href {http://dx.doi.org/10.1093/mnras/stv2304}
  {\path{doi:10.1093/mnras/stv2304}}.

\bibitem{Powell:2014hea}
D.~Powell, T.~Abel, {An exact general remeshing scheme applied to physically
  conservative voxelization}, J. Comput. Phys. 297 (2015) 340--356.
\newblock \href {http://arxiv.org/abs/1412.4941} {\path{arXiv:1412.4941}},
  \href {http://dx.doi.org/10.1016/j.jcp.2015.05.022}
  {\path{doi:10.1016/j.jcp.2015.05.022}}.

\bibitem{Peebles1993}
P.~J.~E. {Peebles}, {Principles of Physical Cosmology}, 1993.

\bibitem{Laine:2011:HSR:2018323.2018337}
S.~Laine, T.~Karras,
  \href{http://doi.acm.org/10.1145/2018323.2018337}{High-performance software
  rasterization on gpus}, in: Proceedings of the ACM SIGGRAPH Symposium on High
  Performance Graphics, HPG '11, ACM, New York, NY, USA, 2011, pp. 79--88.
\newblock \href {http://dx.doi.org/10.1145/2018323.2018337}
  {\path{doi:10.1145/2018323.2018337}}.
\newline\urlprefix\url{http://doi.acm.org/10.1145/2018323.2018337}

\bibitem{Skillman:2014qca}
S.~W. Skillman, M.~S. Warren, M.~J. Turk, R.~H. Wechsler, D.~E. Holz, P.~M.
  Sutter, {Dark Sky Simulations: Early Data Release}\href
  {http://arxiv.org/abs/1407.2600} {\path{arXiv:1407.2600}}.

\end{thebibliography}
